\documentclass[]{elsarticle}
\usepackage[margin=1in]{geometry}
\graphicspath{{Figures/}}
\usepackage{color}

\usepackage{mathbbol} 
\usepackage{amsmath}
\usepackage{wrapfig} 

\usepackage[labelfont=bf]{caption}
\usepackage{accents}
\newlength{\dhatheight}
\newcommand{\doublehat}[1]{%
    \settoheight{\dhatheight}{\ensuremath{\hat{#1}}}%
    \addtolength{\dhatheight}{-0.35ex}%
    \hat{\vphantom{\rule{1pt}{\dhatheight}}%
    \smash{\hat{#1}}}}
\usepackage{hyperref}  
\usepackage{nicematrix}
\newcommand{\bra}[2]{\mbox{$|#2 \rangle $}}
\newcommand{\ket}[2]{\mbox{$ \langle#2 | $}}

\begin{document}
\begin{frontmatter}

\title{Metal Ions based Dynamic Nuclear Polarization: MI-DNP}

\author{Daniel Jard\'{o}n-\'{A}lvarez} 
\author{Michal Leskes*}
\address{Department of Molecular Chemistry and Materials Science, Weizmann Institute of Science, Rehovot, 76100, Israel
*michal.leskes@weizmann.ac.il}

\begin{abstract}
Over the last two decades magic angle spinning dynamic nuclear polarization (MAS DNP) has revolutionized NMR for materials characterization, tackling its main limitation of intrinsically low sensitivity. Progress in theoretical understanding, instrumentation, and sample formulation expanded the range of materials applications and research questions that can benefit from MAS DNP. Currently the most common approach for hyperpolarization under MAS consists in impregnating the sample of interest with a solution containing nitroxide radicals, which upon microwave irradiation serve as exogenous polarizing agents. On the other hand, in metal ion based (MI)-DNP, inorganic materials are doped with paramagnetic metal centres, which then can be used as endogenous polarizing agents. In this work we  give an overview of the electron paramagnetic resonance (EPR) concepts required to characterize the metal ions and discuss the expected changes in the NMR response due to the presence of paramagnetic species. We highlight which properties of the electron spins are beneficial for applications as polarizing agents in DNP and how to recognize them, both from the EPR and NMR data. A theoretical description of the main DNP mechanisms is given, employing a quantum mechanical formalism, and these concepts are used to explain the spin dynamics observed in the DNP experiment. In addition, we highlight the main differences between MI-DNP and the more common approaches in MAS DNP, which use organic radicals as exogenous polarizing source. Finally, we review some applications of metal ions as polarizing agents in general and then focus particularly on research questions in materials science that can benefit from MI-DNP.
\end{abstract}

\end{frontmatter}

Edited by Geoffrey Bodenhausen and David Neuhaus

\tableofcontents

\section{Introduction}
    \subsection{Overview}
    
The main goal of this review is to give a broad overview of the use of paramagnetic metal ions as polarizing agents for dynamic nuclear polarization (DNP) in magic angle spinning (MAS)NMR of inorganic solids. This approach to MAS DNP consists in doping the material of interest with paramagnetic metal ions which then can be used as endogenous sources of polarization for solid state NMR spectroscopy. We will use the term MI-DNP (metal ion based DNP) to refer to this specific procedure. MI-DNP offers unique opportunities for characterization as it provides enhancement of nuclei in the bulk of the material. However, since it is part of the material's structure, the properties of the polarizing agent will be sample specific. In addition, its paramagnetic nature will affect the magnetic resonance properties of the sample itself. Therefore, describing a MI-DNP experiment requires some understanding of EPR, paramagnetic NMR, as well as DNP. In a previous review on metal ions DNP,\cite{CIC_MIDNP_2021} we focused mostly on the DNP process. Here we aim to describe these various subfields of magnetic resonance within a common framework and language. We hope that this will facilitate access for NMR spectroscopists without prior knowledge in EPR or DNP so that they can readily delve into this subject.

\begin{wrapfigure}{l}{0.5\textwidth}
\centering
\includegraphics[scale=0.5]{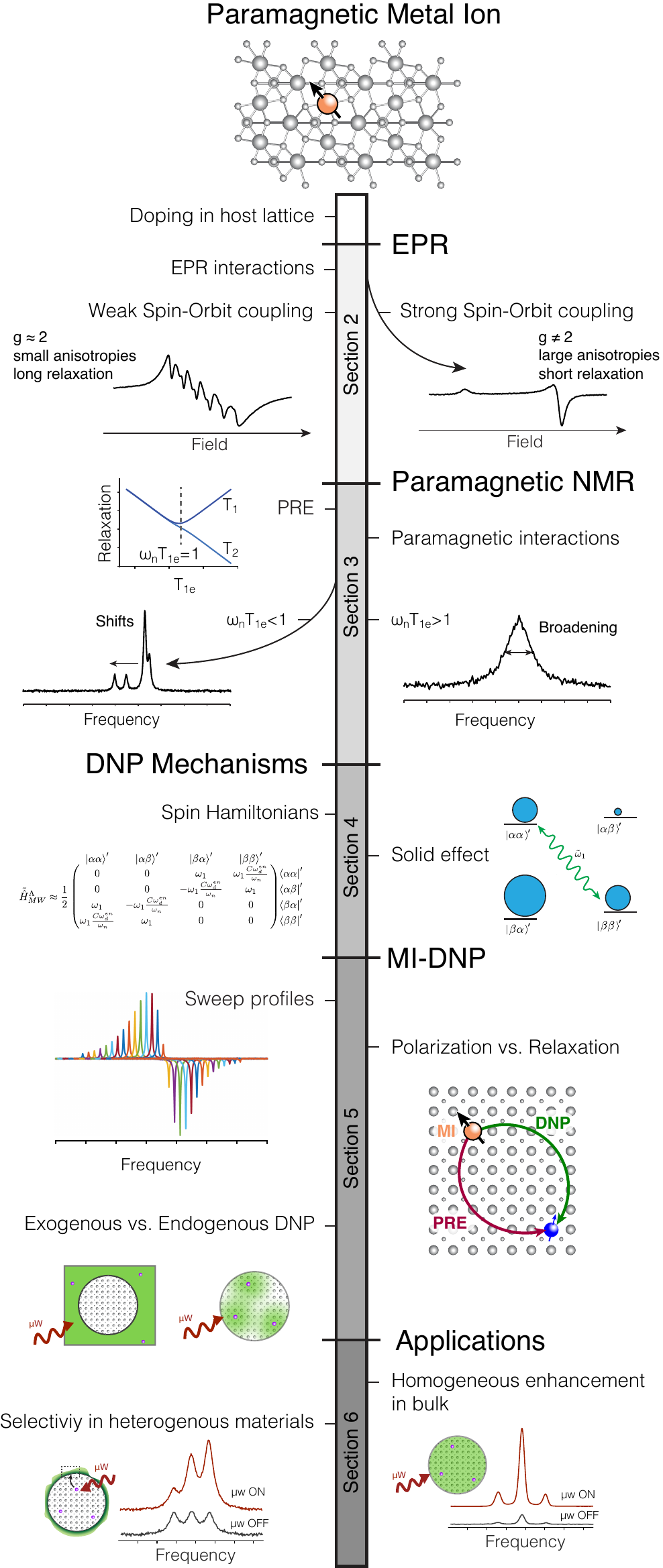}
\caption{Schematic flowchart of the outline of this review, representing the general working scheme in metal ions based DNP.}
\label{fg:flow}
\end{wrapfigure}

Figure~\ref{fg:flow} summarizes schematically the outline of this article. In sections \ref{sec:epr} and \ref{sec:pnmr} we will introduce the basic theoretical concepts of EPR and paramagnetic NMR, by looking at the relevant Hamiltonians and discussing how they will manifest in the respective spectra. In addition, various pertinent relaxation processes are discussed due to their importance in the DNP process. We finish each of these sections with some experimental case studies which enable us to highlight aspects that are critical for MI-DNP. Section~\ref{sec:dnp} covers the fundamentals of  polarization transfer via DNP within a quantum mechanical formalism. By explicitly writing the operators in their matrix representations, together with some basic concepts covered in Appendix~A, we hope that this text can be useful to students interested in learning about DNP in general, independently of the polarizing agent. The main focus will lie on the solid effect mechanism, due to its relevance for MI-DNP, while cross effect and Overhauser effect are only briefly described. 

In section~\ref{sec:MIDNP} we deal with aspects that are more specific to MI-DNP, such as DNP magnetic field sweeps for metal ions with complex EPR spectra or the consequences arising from the fact that the polarizing agents are endogenous. This enables a comparison of MI-DNP with exogenous MAS DNP approaches and emphasizes its unique features. In section~\ref{sec:app}, we draw the historical context from the early use of metal ions in DNP in single crystals, at very low temperatures and low fields, up to their introduction to high field MAS. We finish by summarizing applications of MI-DNP reported to date and highlighting the potential of this technique to assist NMR in materials characterization.

    \subsection{Dynamic nuclear polarization in modern NMR}
The advances of DNP over the past decades have had an immense impact on solid state MAS NMR.\cite{pnmrs_102_120_2017} DNP addresses the most important limitation of NMR, its intrinsically low sensitivity, by providing nuclear polarization beyond the thermal limit set by Boltzmann statistics.\cite{pr_92_411_1953} In DNP the much larger polarization of electron spins is partly transferred to nuclear spins, with the maximum theoretical enhancement of the polarization being given by the ratio $\gamma_e/\gamma_n$ of the gyromagnetic ratios of the electron and the nucleus of interest. Due to the much faster timescale of  electron relaxation compared to nuclear relaxation, a single electron spin can be responsible for hyperpolarizing a large bath of nuclear spins. A DNP experiment requires two basic ingredients beyond regular NMR experiments: First, a source of unpaired electrons, and second, microwave irradiation of the sample. Regarding the latter, the development of  sources capable of emitting microwaves at hundreds of GHz with high power was a critical point for the transition of MAS DNP from low\cite{jmr_52_424_1983} to high magnetic fields\cite{prl_71_3561_1993}. This is key for its current success, as it enables coupling DNP with state-of-the-art NMR spectroscopy. Further developments in DNP instrumentation regarding microwave- sources and their properties, transmission lines and cavities,\cite{jmr_265_77_2016,jmr_289_45_2018,jmr_264_131_2016,jmr_302_43_2019} pulse DNP schemes,\cite{ac_54_11770_2015,jcp_143_054201_2015,sa_5_6909_2019,jacs_144_1513_2022} as well as cryogenic MAS capabilities\cite{jmr_264_99_2016,jmr264_116_2016,jmr_286_1_2018,pccp_23_4919_2021} are crucial for the advances of DNP,\cite{jmr_264_88_2016,pnmrs_102_120_2017,pnmrs_126_1_2021} but are beyond the scope of this review. The second critical point for the success of modern DNP was the development of efficient polarizing agents. These included narrow line radicals\cite{jcp_102_9494_1995} and nitroxide biradicals,\cite{jacs_126_10844_2004,jacs_128_11385_2006} optimized for the cross effect DNP mechanism, as well as the discovery of suitable paramagnetic metal ions.\cite{jacs_133_5648_2011,jacs_136_11716_2014,jacs_141_451_2019} The focus of this review is on the polarizing agents, the properties of the unpaired electron spins that allow prediction of what makes a paramagnetic centre a good polarizing agent; as well as their interactions with nuclear spins. 

\begin{figure*}
\begin{center}
\includegraphics[scale=1]{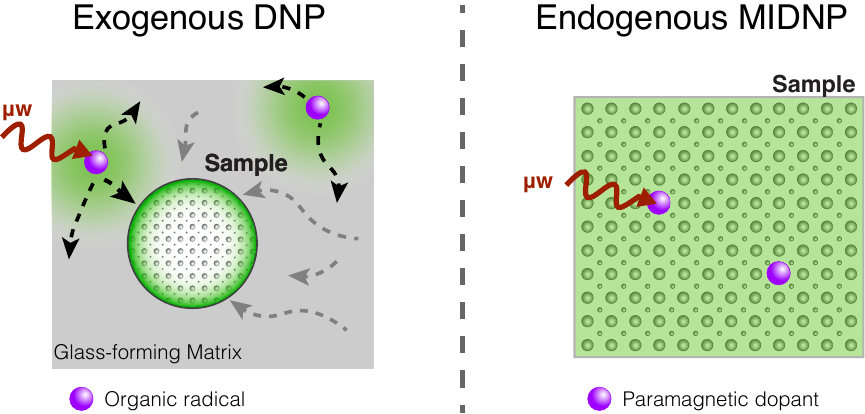}
\end{center}	
\caption{{\bf Comparison of the exogenous DNP formulation and the endogenous metal ion based DNP approach.} In exogenous DNP the polarizing agents are surrounding the sample of interest, while in the endogenous approach they are part of the structure. Upon microwave irradiation, the large polarization from the unpaired electron spins of the polarizing agents is transferred to surrounding nuclear spins. In the exogenous approach the enhancements are mostly limited to surface sites, while in the endogenous approach resonances in the bulk of the sample are enhanced.}
\label{fg:intro}
\end{figure*}

To date, most applications of MAS DNP rely on the use of organic radicals that act as exogenous polarization sources. The radicals are dissolved in a glass-forming solvent. Various different solvent formulations have been optimized, depending on the requirements of the radical and the sample of interest. Subsequently, the sample is soaked or impregnated with the radical-containing solution and rapidly frozen to cryogenic temperatures, commonly $\sim$100~K, to form a glassy matrix (Fig.~(\ref{fg:intro})). Virtually every area of solid-state NMR has benefited from this approach, with applications ranging from proteins and whole cells in structural biology,\cite{cr_122_9738_2022} to catalysts and battery materials in materials science.\cite{ssnmr_101_116_2019,armr_52_25_2022}

The main limitation of the exogenous DNP approach is that the hyperpolarization is generated at the sample surface and has to diffuse into the sample. This inevitably will lead to a polarization gradient, with a decreasing degree of polarization when going deeper into the sample. In protonated samples this might be less critical, since proton spin diffusion is capable of relaying polarization into the core of micron-sized particles,\cite{jacs_134_16899_2012} but for nuclear spins having a low natural abundance and/or gyromagnetic ratio this often limits the enhancements to the first few surface layers.\cite{ac_50_8367_2011}
There are cases where this surface selectivity might actually be desired, as exploited in the DNP SENS approach,\cite{jacs_132_15459_2010,emr_7_93_2018} but often there is a need to increase the sensitivity in the bulk of the material. To enable sensitivity gains in the bulk of large particles with DNP, the use of endogenous polarizing agents is recommended. Among the various types of endogenous agents, the use of paramagnetic metal ion dopants stands out due to its versatility. It is often straight forward to incorporate small amounts of paramagnetic dopants into the material of interest  (Fig.~(\ref{fg:intro})). Historically, the use of paramagnetic dopants or impurities in inorganic lattices has been among the first demonstrations of DNP.\cite{pr_106_165_1957,jpr_19_843_1958} These experiments, however, were performed with single crystals at very low temperatures and low magnetic fields. It was not until very recently that the use of paramagnetic metal ions was demonstrated for MAS DNP applications,\cite{jacs_133_5648_2011,jacs_136_11716_2014,pccp_18_27190_2016} including their use as dopants in inorganic materials, an approach that has come to be known as MI-DNP.\cite{cpc_19_2139_2018,jacs_141_451_2019} The path that lead to these achievements, the underlying spin dynamics and the implications and opportunities of exploiting MI-DNP for materials science are the topics of this review.

\section{EPR}
    \label{sec:epr}
The DNP process requires the presence of an unpaired electron spin which will act as source of polarization. Whether an electron spin can be used as polarizing agent will depend solely on its EPR properties and on its coupling to the environment. Since the EPR properties of a metal ion depend on its structural environment, a thorough EPR characterization of the metal ion in the sample of interest is often essential prior to the DNP experiments. In this section an introduction to some of the basic concepts of EPR interactions (subsection \ref{subsec:epr_interactions}) and relaxation (subsection \ref{subsec:epr_relax}) of paramagnetic metal ions will be given. In subsection \ref{subsec:epr_spec}, a case study is shown to highlight some of the EPR features which can be expected in samples relevant for MI-DNP.

    \subsection{The EPR interactions}
      \label{subsec:epr_interactions}
      \subsubsection{The g-value}
Unlike nuclear spins, electron spins are strongly affected by their surroundings to the point that the Zeeman interaction rarely is the dominant term in the total electron spin Hamiltonian.\cite{pomr} Therefore, in order to find a suitable framework to describe and understand the electron spin resonance properties of a given system, it is important to choose an appropriate set of approximations.\cite{emagres_7_179_2018} 

For transition metals, the ligand field interaction is commonly orders of magnitude larger than any other interaction. Therefore, in the following we shall treat all other interactions as a perturbation to the ligand field interaction.\cite{pnmrs_111_1_2018} A further difference compared to nuclei is that electrons have both a spin angular momentum contribution and an orbital contribution to the magnetic moment.
The coupling of the spin and orbital angular momenta is known as the spin-orbit interaction, which is generally the second largest interaction in the electron spin Hamiltonian.\cite{pnmrs_111_1_2018} 

Interestingly, even though the spin-orbit coupling can be significantly larger than the Zeeman interaction, its contribution to the ground state of the ligand field will be quenched to first order if the ground state is non-degenerate.\cite{pomr,btoepr} The presence of a non-degenerate ground state is not uncommon and can be either due to the effect of the large ligand field or due to half-filled orbital shells.\cite{emagres_7_179_2018}
Nonetheless, contributions from the orbital angular momentum originating from partial mixing of the ground state $\phi_0$ with low excited states $\phi_n$ can have a large effect on the EPR spectrum. These terms can be derived from perturbation theory to second order and can be understood as correction terms to the Zeeman energy of the ligand field ground state due to the admixture of excited states. The resulting effective spin Hamiltonian of the Zeeman interaction becomes:\cite{pnmrs_111_1_2018}
\begin{equation}
\hat{H}_{eZ}=\frac{\mu_B {\bf B_0}}{\hbar}\left(g_e{\bf 1}-2\lambda{\bf \Lambda}\right){\bf\hat{S}},
\label{eq:elecZeeman}
\end{equation}
 with:
\begin{equation}
\Lambda_{ij}=\sum_{n\neq0}\frac{\ket{}{\phi_0}\hat{L}_i\bra{}{\phi_n}\ket{}{\phi_n}\hat{L}_j\bra{}{\phi_0}}{E_n-E_0}
,
\end{equation}
where $i,j=x,y,z$, $\lambda$ is the spin-orbit coupling constant, $B_0$ the external magnetic field, $\mu_B$ the electron Bohr magneton and $g_e\approx2.0023$ the free electron g-factor. The gyromagnetic ratio of the electron is defined from these quantities according to $\gamma_e=-\mu_Bg_e/\hbar$. The Larmor frequency of a free electron is given by $\omega_e=-\gamma_eB_0$. Excited states that are close in energy to the ground state will lead to a deviation of the resonance frequency from $\omega_e$; further, due to the tensorial shape of ${\bf \Lambda}$ it can become anisotropic. Note that throughout the text Hamiltonians will be given in frequency units, following common use in magnetic resonance literature,\cite{sdbonmr,mssnap,popepr} the conversion into energy units being achieved by multiplication with $\hbar$.

In metal ions with half-filled electron shells the energy difference between the ground state and the lowest excited states is particularly large, leading to very low contributions from the orbital momentum.\cite{emagres_7_179_2018,pccp_24_17397_2022} This is the case for high-spin Mn(II), high-spin Fe(III) or Gd(III). As we will see in the following sections, this property makes them ideal candidates for polarizing agents in DNP experiments. For the limiting case of negligible orbital contribution equation~\ref{eq:elecZeeman} simplifies to:
\begin{equation}
\hat{H}_{eZ}=\frac{\mu_Bg_e B_0}{\hbar}\hat{S}_z.
\end{equation}

        \subsubsection{Zero-Field Splitting}
\begin{figure*}
\begin{center}
\includegraphics[scale=1]{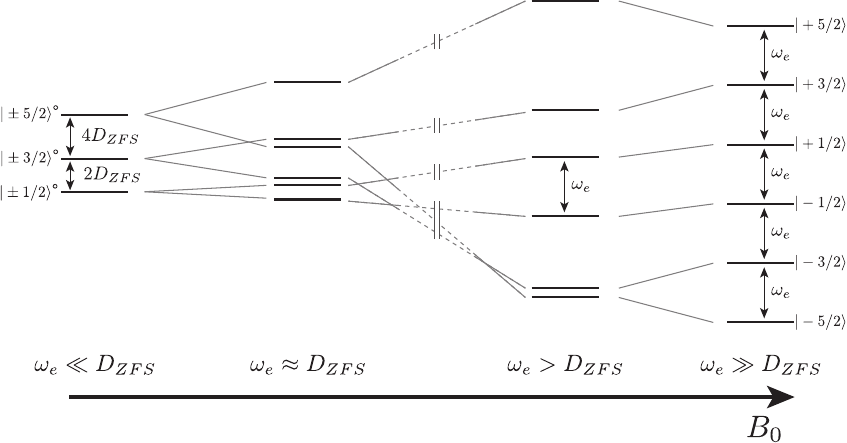}
\end{center}	
\caption{{\bf Effect of the zero field splitting interaction on the EPR transition frequencies.} Energy levels of a spin 5/2 system, such as Mn(II) or Fe(II) ions, considering the Zeeman and the first order ZFS interaction, schematically shown for a single crystallite orientation of axial symmetry. From left to right the ratio $\omega_e/D_{ZFS}$ is growing. Important to note, the basis sets on the left and right are defined with respect to the ZFS interaction frame, as indicated by the circle superscript, and the laboratory frame, respectively. $Left$ side: In the absence of an external magnetic field, the ZFS interaction splits the energy levels into three pairs (Kramers doublets) with two transition frequencies. The degeneracy between the pairs is lifted when subjected to an external magnetic field due to the Zeeman interaction. $Right$ side: In the presence of Zeeman only, the 6 energy levels are equally spaced by the transition frequency $\omega_e$. In the presence of ZFS there will be five transition frequencies: one cetnral transition (CT) between $+1/2\leftrightarrow-1/2$, two inner- $\pm1/2\leftrightarrow\pm3/2$ and two outer $\pm3/2\leftrightarrow\pm5/2$ satellite transition (ST). To first order, the CT is unaffected by the ZFS. }
\label{fg:ZFS}
\end{figure*}

In metal ions with more than one unpaired electron the total spin number will be $S=n/2$, where $n$ is the number of unpaired spins. 
Interactions among the spins, or differential interactions of the spins with the orbital angular momentum, give rise to an additional interaction called the zero-field splitting (ZFS).\cite{emr_207_2017} The ZFS Hamiltonian is given by:
\begin{equation}
\hat{H}_{ZFS}^{\circ}=D_{ZFS}\left[\hat{S}_z^{\circ2}-\frac{1}{3}S(S+1)\right]+E_{ZFS}\left[\hat{S}_x^{\circ2}-\hat{S}_y^{\circ2}\right],
\end{equation}
with $D_{ZFS}= 3D_z^{\circ}/2$ and $E_{ZFS} = (D_x^{\circ}- D_y^{\circ})/2$ defined from the principle axis frame components of the interaction tensor $\boldsymbol{D}^{\circ}$.\cite{popepr} The circle superscript indicates that the operators are defined in the principle axis frame of the ZFS interaction. 
For an axial symmetric case ($E=0$) it can readily be seen that in the absence of an external magnetic field the energy levels $\bra{}{\pm m_S}$ will remain degenerate, since $\hat{S}_z^2\bra{}{\pm m_S}=m_S^2\bra{}{\pm m_S}$. These are called Kramer doublets. The splitting due to the ZFS interaction only is shown in Fig.~\ref{fg:ZFS} for a spin 5/2. While for rhombic symmetry ($E>0$) the second term in the equation can lift the degeneracy, this effect is negligible in half-integer spins.\cite{emr_207_2017}

Often, in the presence of a large external magnetic field, it is more useful to express the Hamiltonian in the laboratory frame (the frame defined by the eigenbasis of the Zeeman interaction), following the notation of Ref.~\cite{bmr_73_1984}, one can rewrite the Hamiltonian according to:
\begin{equation}
\hat{H}_{ZFS}
=D_0\left[3\hat{S}_z^2-S(S+1)\right]
+D_{+1}\left[\hat{S}_+\hat{S}_z+\hat{S}_z\hat{S}_+\right]
+D_{-1}\left[\hat{S}_-\hat{S}_z+\hat{S}_z\hat{S}_-\right]
+D_{+2}\left[\hat{S}_+^2\right]
+D_{-2}\left[\hat{S}_-^2\right]
,
\label{eq:ZFS_lab}
\end{equation}
with:
\begin{equation}
\begin{split}
D_0=&
\frac{D_{ZFS}}{3}\left(3\cos^2\theta-1\right)+\frac{E_{ZFS}}{2}\sin^2\theta\cos2\phi,\\
D_{\pm1}=&\left(\frac{1}{4}\sin2\theta\right)\left(-D_{ZFS}+E_{ZFS}\cos2\phi\right)\pm\frac{i}{2}E_{ZFS}\sin\theta\sin2\phi,\\
D_{\pm2}=&\frac{1}{4}\left[D_{ZFS}\sin^2\theta+E_{ZFS}\cos2\phi\left(1+\cos^2\theta\right)\right]\pm\frac{i}{2}E_{ZFS}\cos\theta\sin2\phi
,
\end{split}
\end{equation}
here $\theta$ and $\phi$ are the two Euler angles required to describe the transformation from the principle axis frame to the laboratory frame. By looking at the spin operators in equation~(\ref{eq:ZFS_lab}) it is evident that there are diagonal and off-diagonal terms, i.e., terms that do and do not commute with the Zeeman interaction. To find the energy levels in the presence of ZFS one would need to diagonalize this Hamiltonian. To avoid this tedious step, when the Zeeman interaction is significantly larger than the ZFS, it is possible to use a perturbation approach instead. The first two correction terms to the Zeeman energy are:
\begin{equation}
E_{m_S}^{(1)}=D_0\left(3m_S^2-S(S+1)\right),
\end{equation}
and
\begin{equation}
E_{m_S}^{(2)}=
\frac{1}{\omega_e}\left[
D_{+1}D_{-1}\left(c_{+1}-c_{-1}\right)
+\frac{1}{2}D_{+2}D_{-2}\left(c_{+2}-c_{-2}\right)
\right]
,
\end{equation}
where
\begin{equation}
\begin{split}
c_{\pm1}=&\left(2m_S\mp1\right)^2
\left[S(S+1)-m_S\left(m_S\mp1\right)\right],\\
c_{\pm2}=&\left[S(S+1)-m_S\left(m_S\mp1\right)\right]
\left[S(S+1)-\left(m_S\mp1\right)\left(m_S\mp2\right)\right].
\end{split}
\end{equation}
These equations show that, to first order, energy shifts of levels of the same $|m_S|$ are equal. Therefore, the central transition (CT, $+1/2\leftrightarrow-1/2$) is not affected by the ZFS interaction to first order, as can be seen in Fig.~(\ref{fg:ZFS}). The angular dependence leads to a powder pattern with a width of the order of the interaction itself for all allowed transitions except the CT, which is therefore  significantly narrower  (the basic selection rule of allowed transitions in magnetic resonance is $\Delta m=1$). The broadening of the CT from second order contributions is scaled by the electron Larmor frequency and thus becomes less prominent with increasing external magnetic field. A further important consequence of the presence of a large ZFS is that it can lead to selective excitation of individual transitions. Without going into the derivations, the relevant outcome of this effect is that for the selective excitation of the CT, the nutation frequency $\omega_1$ of the CT becomes $\omega_{1,CT}=(I+1/2)\omega_1$ (for $\omega_{ZFS}\gg\omega_1$),\cite{cpl_174_595_1990} in analogy to the nutation behaviour of the CT in quadrupolar nuclei under selective excitation.\cite{prb_28_6567_1983,cpl_111_171_1984}

        \subsubsection{Hyperfine coupling}
        
The coupling between an electronic and a nuclear spin is called hyperfine interaction (HF). One differentiates between two types of  HF interactions. The Fermi contact (FC) interaction describes the interaction due to part of the electron density being at the same position as the nucleus. In the absence of an orbital angular momentum the FC interaction is isotropic. The dipole-dipole interaction describes the through-space dipolar coupling. 

\begin{figure*}
\begin{center}
\includegraphics[scale=1]{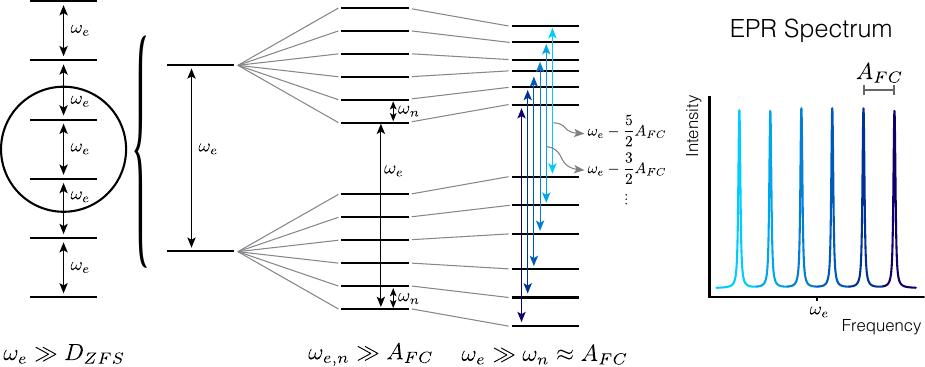}
\end{center}	
\caption{{\bf Effect of the Fermi contact hyperfine interaction on the EPR transition frequencies.} High spin Mn(II) has a spin S=5/2. In addition, the isotope $^{155}$Mn, which is 100\% abundant, has a spin 5/2 nucleus. Due to the Fermi contact interaction between electronic and nuclear spin each electron spin transition (here only shown for CT) splits into 6 transitions, centred at $\omega_e$ and spaced by $A_{FC}$.}
\label{fg:HFC}
\end{figure*}

The Hamiltonian of the FC interaction for the isotropic case is given by:
\begin{equation}
\hat{H}_{FC}=A_{FC}{\bf \hat{I}\cdot \hat{S}}
,
\label{eq:FC}
\end{equation}
where the dot product means:
\begin{equation}
{\bf \hat{I}\cdot \hat{S}}=\hat{I}_x\hat{S}_x+\hat{I}_y\hat{S}_y+\hat{I}_z\hat{S}_z,
\end{equation}
and 
\begin{equation}
A_{FC}=-\frac{2\mu_0}{3}\gamma_I\gamma_S\rho,
\end{equation}
where $\mu_0$ is the vacuum magnetic permeability and $\rho$ is the spin density of the electron at the position of the nucleus.\cite{emr_271_2017} The secular term ($\hat{I}_z\hat{S}_z$) will split the EPR signal into (2I+1) peaks. The contribution from the hyperfine coupling to any transition $\bra{}{m_S,m_I}\leftrightarrow\bra{}{m_S-1, m_I}$ can be obtained from:
\begin{equation}
\Delta E_{FC}=\hbar A_{FC}\hat{I}_z\hat{S}_z\bra{}{m_S-1,m_I}-\hbar A_{FC}\hat{I}_z\hat{S}_z\bra{}{m_S,m_I}=-m_I\hbar A_{FC}.
\end{equation}
For a spin 1/2 electron coupled to a spin 1/2 nucleus this leads to a doublet at $\omega_e\pm\frac{1}{2}A_{FC}$. Analogously, for higher nuclear spins, all peaks will be separated by $A_{FC}$ (see Fig.~(\ref{fg:HFC})). Note that, while the splitting is independent of the nuclear Larmor frequency, $A_{FC}$ is proportional to the nuclear gyromagnetic ratio and therefore, if there are several isotopes of the nucleus of a paramagnetic metal ion, the relative magnitude of the Fermi contact hyperfine splittings will reflect the ratio in $\gamma_n$.

The non-secular terms can be treated in terms of perturbation theory.\cite{bmr_73_1984} Their effect can lead to further complications in the EPR spectrum through the appearance of formally forbidden transitions, but this will not be further discussed here.\\

The Hamiltonian for the dipole-dipole coupling in the point-dipole approximation, where the magnetic moment of the electron is assumed to be localized in space, is given by:
\begin{equation}
\hat{H}_{dd}=\omega_d^{en}\left({\bf \hat{I}\cdot \hat{S}}-3\frac{\left({\bf \hat{I}\cdot r}\right)\left({\bf \hat{S}\cdot r}\right)}{r^2}\right)=\omega_d^{en}\left( {\bf \hat{I} D \hat{S}}\right)
,
\label{eq:dip}
\end{equation}
where ${\bf D}$ is the dipolar coupling tensor
\begin{equation}
{\bf D}=
 \begin{pmatrix} 
 \left(1-3\sin^2\theta \cos^2\phi\right) & -3\sin^2\theta \sin\phi \cos\phi & -3\sin\theta \cos\theta  \cos\phi\\
 -3\sin^2\theta \sin\phi \cos\phi &\left(1-3\sin^2\theta \sin^2\phi\right) & -3\sin\theta \cos\theta  \sin\phi\\
 -3\sin\theta \cos\theta \cos\phi  & -3\sin\theta \cos\theta  \sin\phi &\left(1-3\cos^2\theta\right)
 \end{pmatrix},
\label{eq:Dtensor}
\end{equation}
with $\theta$ and $\phi$, the spherical polar angles describing the orientation of the vector connecting both spins with respect to the external magnetic field, and 
\begin{equation}
\omega_d^{en}=-\frac{\mu_0}{4\pi}\frac{\gamma_I\gamma_S\hbar}{r^3}.
\label{eq:wd}
\end{equation}

In principle, through-space dipolar interactions between a paramagnetic metal ion and nuclei in the surrounding diamagnetic lattice can provide unique structural insights. Unfortunately, these interactions can be difficult to detect with CW EPR methods, due to their anisotropic nature and generally small size compared to other sources of EPR line broadening. However, if relaxation times are sufficiently long, pulsed EPR methods can allow one to disentangle dipolar couplings from other interactions. Such pulsed EPR methods are described in detail in  Refs.\cite{popepr,eprsfm}. We will  delve further into the effects of dipolar interactions in subsequent sections since this interaction is essential for MI-DNP.

    \subsection{Electron spin relaxation}
       \label{subsec:epr_relax}
In magnetic resonance there are two fundamental relaxation processes: longitudinal relaxation, which represents the return of the populations of the eigenstates to Boltzmann's equilibrium after a perturbation; and transverse relaxation, which results from the loss of coherence excited by a pulse. These relaxation processes are characterized by the relaxation times $T_1$ and $T_2$, respectively. To differentiate between nuclear and electron spin relaxation times, we will use the subscript $e$ for the latter throughout the text, $T_{1e}$ and $T_{2e}$. A phenomenological description of the two relaxation processes is given in the appendix.

As previously mentioned, the interactions of electron spins with the surroundings can be much larger than for nuclear spins. In addition, there are various different mechanisms that couple the electron spins with the lattice to drive longitudinal relaxation. Consequently, theoretical treatments of electron spin relaxation are only applicable under specific conditions. For instance, the validity of Redfield's theory is mostly limited to cases with low orbital contributions and large mobility, for example in highly symmetric metal complexes in solution. Relaxation is then driven by stochastic modulations of one of the spin interactions.
On the other hand, mechanisms in the solid state can involve coupling of the orbital moment with the lattice via phonon processes.\cite{nmrpm}

Unfortunately, apart from some limiting cases, predicting or even discerning which electron relaxation mechanism dominates in a given system can be an extremely challenging endeavour. In an effort to summarize various mechanisms, G.R. Eaton and S.S. Eaton give the following equation for the longitudinal relaxation rate $R_{1e}$, emphasizing their distinct temperature dependence:\cite{jmr_139_165_1999,emr_1543_2016}
\begin{equation}
R_{1e}=A_{dir}B_0^4T
+A_{Ram}\left(\frac{T}{\Theta_D}\right)^9J_8\left({\Theta_D}/{T}\right)
+A_{Orb}\left(\frac{\Delta_{Orb}^3}{\exp\left(\Delta_{Orb}/T\right)-1}\right)
+A_{therm}J(\omega_e)
.
\end{equation}
The four terms represent the direct, Raman, Orbach and thermally activated relaxation mechanisms. In addition, molecular species can also be relaxed through local-mode processes. The terms $A_n$ are the respective coefficients, $B_0$ is the external magnetic field, $\Theta_D$ the material-specific Debye temperature, $\Delta_{Orb}$ the energy difference between the ground and excited states,  and $J_8\left({\Theta_D}/{T}\right)$ is the transport integral. In the first three mechanisms, transitions are caused by an energy match between the electron spin transition and one lattice phonon (direct relaxation mechanism) or two lattice phonons (Raman and Orbach mechanisms). Finally, the thermal activated process includes relaxation described by Redfield's theory, where $J(\omega_e)$ is the familiar spectral density function at the electron Larmor frequency. Here, any stochastic motion modulating the local magnetic field at the electron spin will lead to relaxation and will be most efficient when the correlation time of the motion is equal to the inverse of the electron Larmor frequency.\cite{eproti,nmrpm,rops} 
An important aspect to note is the markedly distinct temperature dependence of the different mechanisms. This in principle could allow one to differentiate among the various relaxation processes if sufficient data is available.

Bertini et al. estimated typical electron relaxation times for paramagnetic metal ions at high magnetic fields.\cite{nmrpm} While these values are for metal ions in solution and at room temperature it is still instructive to study them for the purpose of MI-DNP. For instance rare earth metal ions have generally shorter relaxation times compared to d-block transition metals, with values very roughly around 10$^{-13}$~s for the former and between 10$^{-10}$ to 10$^{-11}$~s for the later. This difference originates from the fact that the f-orbitals are less exposed to the ligands and, therefore, the energy gap caused by the ligand field is smaller, leading to a larger admixture of orbital momentum. More important for our purposes is to look at the outliers and realize that whenever the orbital shells are exactly half filled, significantly longer relaxation times can be expected, of up to 10$^{-8}$~s. As we will see later, even longer relaxation times may be desirable for DNP, often requiring experiments to be performed at cryogenic temperatures. Again, the explanation for this is the efficient quenching of the orbital momentum in metal centres with half-filled shells.

   \subsection{The EPR spectrum}
          \label{subsec:epr_spec}

In continuous wave (CW) EPR experiments, the magnetic field is varied while irradiating at a fixed frequency. In addition to the sweep, the magnetic field is also modulated at any point. The response signal thus is the derivative of the absorption, which is generally what is plotted in a CW EPR spectrum. The absorption spectrum can be obtained by integration. The intensity of the signal will depend on the microwave power $P$ and the relaxation parameters of the spins, and is given by:\cite{bba_537_255_1978,jmr_322_106875_2021}
\begin{equation}
S_{pp}=\frac{A\sqrt{P}}{(1+\gamma_e^2C^2T_{1e}T_{2e}P)^c}
,
\end{equation}
where $A$ is a scaling factor related to the number of spins in the resonator, $C$ the conversion efficiency of the instrument, and $c$ an empirical exponent. At low microwave power the signal intensity increases linearly with the square root of the power (note that the microwave magnetic field $B_1$ is proportional to the square root of the power $\sqrt{P}$). On the other hand, excessive power will lead to saturation of the spin transition and lowering of the signal intensity. The parameter $c$ in the equation is a measure of the homogeneity of the signal,\cite{bba_537_255_1978,jcp_70_3300_1979} and can take values between 1.5 (fully homogeneous limit) and 0.5 (inhomogeneous case). In homogeneous lines the signal intensity decreases to zero with high power, while in inhomogeneous cases it levels off at a plateau of value larger than zero.\cite{eprJWS} Measuring the signal intensity as a function of the microwave power can be a simple method to gain useful information about the relaxation properties of the electron spins in the sample. Unfortunately, this equation does not differentiate between $T_{1e}$ and $T_{2e}$, however, for homogeneously broadened lines, $T_{2e}$ might be extracted from the width of the Lorentzian-shaped signal.\\

\begin{figure*}
\begin{center}
\includegraphics[scale=1]{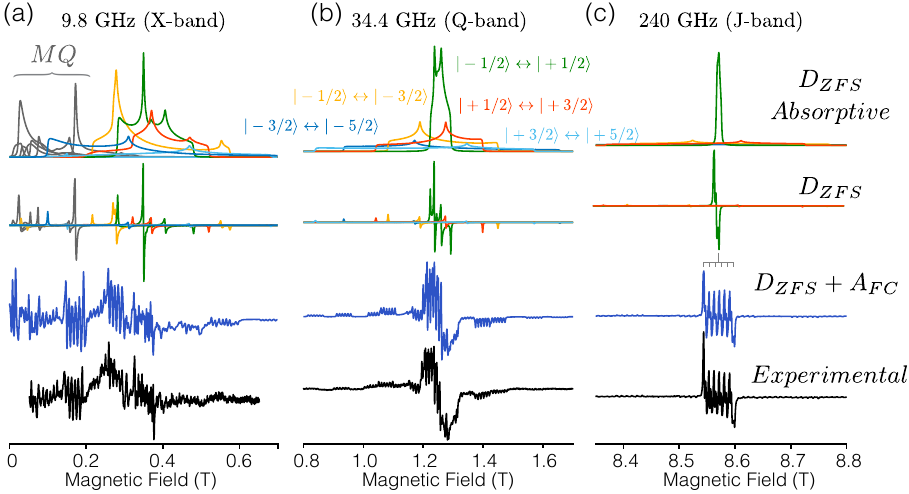}
\end{center}	
\caption{{\bf Example of experimental Mn(II) EPR spectra.} Experimental (bottom) continuous wave EPR spectra of LiMg$_{1-x}$Mn$_x$PO$_4$, with $x=0.005$, at microwave irradiation frequencies of 9.8 (a), 34.3 (b) and 240~GHz (c). Above, respective simulated spectra. Simulations from bottom to top: first, in blue, best fit spectrum, above is the spectrum without hyperfine interaction in first derivative mode and on the very top in absorptive mode.  The simulation parameters used are: $g_{iso}=1.999$ (in (c); (a) and (b) required a small additional shift), $^{55}$Mn hyperfine coupling $A_{FC}=259\text{ MHz}$ and ZFS interaction $D_{ZFS}=2810\text{ MHz}$ and $E_{ZFS}=531\text{ MHz}$ with a D-strain of 30~MHz. Additional line broadening was added for best agreement with the experimental results. Simulations were done with  the EASYSPIN simulation package.\cite{jmr_178_42_2006} Adapted with permission from Ref.~\cite{jmr_336_107143_2022}.}
\label{fg:EPR}
\end{figure*}

Fig.~(\ref{fg:EPR}) shows the CW EPR spectra of LiMg$_{1-x}$Mn$_x$PO$_4$ with $x=0.005$ obtained at various magnetic fields. In the following we will rationalize the observed spectra and  discuss how the results can be informative to evaluate whether this system is appropriate for MI-DNP applications.
The manganese ions substitute magnesium in the olivine structure where they occupy octahedral sites. High-spin Mn(II) has five unpaired electrons, thus each of the five d-orbitals is half occupied. At this electronic configuration the energy gap between the electronic ground state and the lowest excited state is large, therefore,  orbital angular momentum contributions are very small.\cite{pccp_24_17397_2022} This leads to a g-value close to that of the free electron (here, $g=1.999$), small g-anisotropy (not visible here) and long electron relaxation times (in this case, $T_{1e}$ is in the order of few $\mu$s at 100~K and 9.4~T -- determined by other measurements\cite{jmr_336_107143_2022}). As we will see later, all three aspects of the quenching of the orbital angular momentum may be beneficial for polarizing agents in DNP. The g-value corresponds to resonance frequency within the range of commercially available DNP instrumentation, and the small anisotropy and long relaxation time facilitate a high saturation efficiency.

Manganese has one stable nuclear isotope with 100\% natural abundance, $^{55}$Mn, which has a nuclear spin $I=5/2$ that splits the EPR transitions of Mn(II) into sextets. The hyperfine coupling due to the Fermi contact interaction between the manganese electrons and nuclear spins in this olivine is $A_{FC}=259\text{ MHz}$. Generally, the value $A_{FC}$ for manganese can vary between about 120 and 300~MHz and depends on the coordination symmetry and the covalent character of bonds with coordinating ions.\cite{eproti,jpc_71_51_1967} The hyperfine coupling is independent of the external magnetic field.

Associated with the high-spin number ($S=5/2$) is a ZFS interaction. In the olivine structure the ZFS parameters are $D_{ZFS}=2810\text{ MHz}$ and $E_{ZFS}=531\text{ MHz}$. The powder patterns of the five single-quantum transitions are best recognized when plotted in absorptive mode, as on top of Fig.~(\ref{fg:EPR}). Already at Q-band (34.4~GHz) the central transition (CT) powder pattern, which in the high field approximation is affected by the ZFS only to second order, can be easily differentiated from the broader satellite transitions (ST). As the second order effect scales inversely with the Larmor frequency the CT becomes significantly narrower at J-band (240~GHz), and its intensity much larger compared to the broad ST powder patterns. On the other hand, CT and ST strongly overlap at X-band (9.8~GHz) where, in addition, signals corresponding to formally forbidden multiple quantum transitions become significant at such low field.

The combined effect of ZFS and hyperfine interactions can lead to very complicated spectra, as in the X-band spectrum, where it would be challenging to disentangle the various contributions without knowledge that can be only obtained at higher fields.
The narrowing of the CT can be best appreciated by comparing the width of the powder pattern with the frequency range of the hyperfine manifold.
In terms of high field MAS DNP, it is important to note that the ZFS is scaled sufficiently at fields of interest to fully resolve the sextet of the CT without significant overlap. But it should be noted that the ZFS can be a source of relaxation, so that a large ZFS might become detrimental. A recent study showed a systematic decrease of the DNP efficiency with increasing ZFS, in a quadratic dependence, although the authors attributed this effect mainly to line broadening of the CT.\cite{jpcc_126_11310_2022}
Nonetheless, in agreement with our discussion of low orbit contributions, Mn(II) in this environment appears to be a good candidate to be a polarizing agent, which was confirmed with signal enhancements of more than factor 10 for both $^6$Li and $^7$Li.\cite{jmr_336_107143_2022}

\section{Paramagnetic NMR}
    \label{sec:pnmr}

The field of paramagnetic NMR is extremely rich and broad. Since the magnetic moment from paramagnetic centres is orders of magnitude larger than that from nuclear spins, the presence of even small numbers of unpaired electrons (in concentration as low as a few 100 ppm) can significantly alter the NMR properties of otherwise diamagnetic samples. Effects arising from the presence of paramagnetic centres are enhanced relaxation, line broadening, and frequency shifts. The nature and extent of these effects will depend on the proximity between the paramagnetic species and the nuclei, as well as on the EPR properties of the former. This interplay between paramagnetic centres and surrounding nuclei can be highly sensitive to structural properties and is therefore exploited in many applications of NMR, from biology to materials science.\cite{cr_104_4493_2004,acr_40_206_2007,cr_111_530_2011,ssnmr_43_1_2012,pnmrs_111_1_2018,pccp_24_17397_2022}

In addition to the spin magnetic moment, paramagnetic species can have contributions to the magnetic moment from the orbital. As for the EPR spectrum, the presence of orbital magnetic moment gives rise to additional interactions. For instance, its anisotropic nature can result in an anisotropic susceptibility tensor, which can lead to an isotropic contribution of the through-space dipolar coupling, the so called pseudocontact shift.\cite{jcp_29_1361_1958,jmr_2_286_1970,nmrpm} However, as previously mentioned, the presence of strong orbital contributions is detrimental to the MI-DNP process, therefore, we will focus in this section mostly on the effect of spin-only paramagnetic centres on the surrounding nuclei. For further simplicity, in this section we will assume that the Zeeman interaction is much larger than any other spin interaction, including the ZFS. Here we just intend to give a brief introduction to some of the relevant concepts, a much more detailed treatment, which goes beyond these simplifying assumptions can be found in a book by Bertini et al.\cite{nmrpm} and a review by Pell et al.\cite{pnmrs_111_1_2018}

The Hamiltonians of the relevant electron-nuclear interactions, required for a theoretical description of the spin system and dynamics, are discussed in subsection \ref{subsec:pnmr_interact}. Next, in subsection \ref{subsec:pre}, the paramagnetic relaxation enhancement (PRE) effect in rigid solids is described and a discussion of its dependence on the electron spin relaxation, the concentration of paramagnetic centres as well as on the efficiency of spin diffusion among the nuclear spins is given. In the limit of fast electron relaxation, the paramagnetic NMR spectrum is conveniently described in terms of the magnetic susceptibility. The relevant concepts will be given subsection \ref{subsec:susceptibility}. In the end of the section some examples are given to summarize what changes in the NMR spectrum can be expected when doping a sample with paramagnetic metal ions, and how we can make use of this knowledge to understand or predict MI-DNP applications.

    \subsection{Paramagnetic NMR interactions}
    \label{subsec:pnmr_interact}

    \subsubsection{Fermi contact interaction}
    
We have introduced the Fermi contact (FC) interaction in section~\ref{subsec:epr_interactions}. In this section it will be useful to rewrite the Hamiltonian given in equation~(\ref{eq:FC}) as:
\begin{equation}
\hat{H}_{FC}=A_{FC}\left[\frac{1}{2}\left(\hat{I}_+\hat{S}_-+\hat{I}_-\hat{S}_+\right)+\hat{I}_z\hat{S}_z\right].
\end{equation}

When looking at the Fermi contact interaction from the point of view of the electron spin of the metal ion, we were mostly concerned with the interaction between electron and nuclear spins of the same atom, as this is the strongest interaction. From an NMR point of view this interaction however is likely too strong to enable the measurement of a nuclear spin in a paramagnetic centre. However, electron spin density from the paramagnetic ion can have a non-vanishing contribution to surrounding nuclei through chemical bonds and result in very significant paramagnetic interactions affecting the nuclear spins. Characterizing this interaction can be a valuable tool for studying the chemical properties  of the paramagnetic centre in the host lattice.\cite{cr_104_4493_2004}

In terms of DNP, the FC interaction is the driving interaction in Overhauser DNP from conducting electrons in metals.\cite{pr_92_411_1953,nc_11_2224_2020} In MI-DNP, however, this interaction has not played a significant role in the polarization transfer so far. This has two main reasons: first, MI-DNP requires a very low concentration of metal ions, thus nuclei experiencing FC interaction (typically limited to a few bonds from the metal ion) will be extremely scarce. And second, due to the interaction itself the energy levels of the coupled nuclei will be shifted and the ability to participate in the spin diffusion processes will be severely affected through a mismatch in the energy matching condition with neighbouring nuclei.\cite{pr_119_79_1960}

    \subsubsection{Through-space dipolar couplings}

Unlike the Fermi contact interaction, the through-space dipolar coupling between electron and nuclear spins can potentially affect the NMR properties of nuclei as far as several tens of nanometers away from the paramagnetic centre. The dipolar coupling  $\omega_d^{en}$, defined in equation~(\ref{eq:wd}), decreases with $r^{-3}$. Particularly in the context of MI-DNP in inorganic materials, this interaction will be the driving force of both the relaxation and hyperpolarization of the nuclear spin bath. For the following it will be convenient to write the Hamiltonian given equation~(\ref{eq:dip}) in terms of the so-called dipolar alphabet:
\begin{equation}
\hat{H}_{dd}^{en}=\omega_d^{en}\left[A\hat{I}_z\hat{S}_z
+B\left(\hat{I}_+\hat{S}_-+\hat{I}_-\hat{S}_+
\right)
+C\left(\hat{I}_z\hat{S}_++\hat{I}_+\hat{S}_z
\right)
+D\left(\hat{I}_z\hat{S}_-+\hat{I}_-\hat{S}_z
\right)
+E\hat{I}_+\hat{S}_+
+F\hat{I}_-\hat{S}_-
\right],
\label{eq:dipalph}
\end{equation}
with
\begin{equation}
\begin{split}
A=&3\cos^2\theta-1,\\
B=&-\frac{1}{4}\left(3\cos^2\theta-1\right),\\
C=&\frac{3}{2}\sin\theta\cos\theta\exp(-i\phi),\\
D=&\frac{3}{2}\sin\theta\cos\theta\exp(+i\phi),\\
E=&\frac{3}{4}\sin^2\theta\exp(-2i\phi),\\
F=&\frac{3}{4}\sin^2\theta\exp(+2i\phi).
\end{split}
\end{equation}
Note that in these definitions the spin operators are not included in the terms $A$ to $F$. We should remember that this Hamiltonian is defined within the point-dipole approximation.

    \subsection{Paramagnetic relaxation enhancement}
        \label{subsec:pre}
A phenomenological description of the longitudinal and transverse relaxation processes is given in the Appendix. Here we are concerned with interactions and motions that drive relaxation processes. In NMR, relaxation is mediated by random fluctuations of local magnetic fields.\cite{pomr,nsril,ponm} In general, atomic and molecular motions will alter the magnitude of the spin interactions in a stochastic manner, leading to relaxation. The relaxation rates will depend on both the timescales of the fluctuations as well as on the magnitudes of the changes of the interactions. Longitudinal relaxation is mediated by fluctuations at the Larmor frequency, while transverse relaxation is also mediated by slow fluctuations. In rigid solids, motions often fail to cause relaxation efficiently, either because they do not happen on the {\it right} timescale, or because the magnitude of the induced changes in the interactions are not sufficient, or a combination of the two. When introducing paramagnetic metal ions into an otherwise diamagnetic material, the paramagnetic centres will act as an efficient source of relaxation, an effect known as paramagnetic relaxation enhancement (PRE). The through-space dipolar coupling between the nuclear spin and the magnetic dipole moment of the electron spins will be modulated by the fluctuations of the electron spin itself.\cite{pr_166_279_1968} The PRE effect will depend on the magnitude of the magnetic moment of the electron spins\cite{ssnmr_5_151_1995} and on the correlation time of the fluctuations of its longitudinal and transverse components, $\tau_{1e}$  and $\tau_{2e}$.  The longitudinal and transverse PRE relaxation rates are given by:\cite{nmrpm}

\begin{equation}
R_1=
\frac{2}{3}\left(\omega_d^{en}\right)^2S(S+1)
\left(4A^2
\frac{\tau_{2e}}{1+\tau_{2e}^2(\omega_e+\omega_n)^2}
+2CD
\frac{\tau_{1e}}{1+\tau_{1e}^2\omega_n^2}
+4FE
\frac{\tau_{2e}}{1+\tau_{2e}^2(\omega_e-\omega_n)^2}
 \right),
\label{eq:T1PRE}
\end{equation}
and
\begin{equation}
R_2= 
\frac{R_1}{2}+
\frac{2}{3}\left(\omega_d^{en}\right)^2S(S+1)
\left(\frac{B^2}{2}
{\tau_{1e}}
+2CD
\frac{\tau_{2e}}{1+\tau_{2e}^2\omega_e^2}
 \right).
\label{eq:T2PRE}
 \end{equation}
 
The electron Larmor frequency is orders of magnitude larger than the nuclear Larmor frequency. Therefore, for  $\frac{\tau_{2e}}{1+\tau_{2e}^2\omega_e^2}\ll\frac{\tau_{1e}}{1+\tau_{1e}^2\omega_n^2}$ (which is valid for any reasonable ratio of $\tau_{1e}/\tau_{2e}$) and $\tau_{2e}\omega_e>1$ these expressions can be further simplified to:
 \begin{equation}
R_1=
\frac{2}{3}\left(\omega_d^{en}\right)^2S(S+1)
\left(
2CD
\frac{\tau_{1e}}{1+\tau_{1e}^2\omega_n^2}
 \right),
\label{eq:T1PRE_c}
\end{equation}
and
\begin{equation}
R_2= 
\frac{R_1}{2}+
\frac{2}{3}\left(\omega_d^{en}\right)^2S(S+1)
\left(\frac{B^2}{2}
{\tau_{1e}}
 \right).
\label{eq:T2PRE_c}
 \end{equation}

Often the electron relaxation is the main source of  fluctuation, in that case one can replace the correlation times $\tau_{1e}$  and $\tau_{2e}$ by $T_{1e}$ and $T_{2e}$. However, we note that in principle it is also possible that coherent  processes, such as dipolar interactions between electron spins,  contribute to the fluctuations leading to nuclear relaxation.\cite{jpcc_122_1932_2018}

\begin{figure*}
\begin{center}
\includegraphics[scale=1]{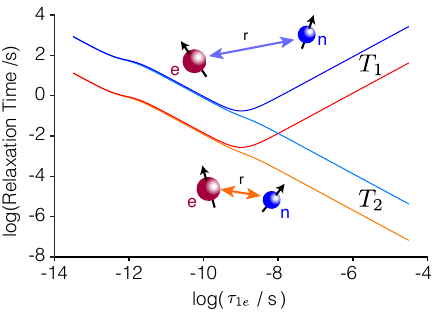}
\end{center}	
\caption{{\bf Theoretical description of nuclear paramagnetic relaxation enhancement.} Nuclear $T_1$ and $T_2$ relaxation times calculated using equations~(\ref{eq:T1PRE}) and (\ref{eq:T2PRE}) as a function of the correlation time $\tau_{1e}$ and with $\tau_{2e}=\tau_{1e}$ for a $^7$Li nucleus coupled to a spin 1/2 electron at distances of 5 and 10~{\AA}, shown in red and blue, respectively. The minimum of $T_1$ is independent of the distance at $\omega_n\tau_{1e}=1$, the step at short $\tau_{1e}$ corresponds to the condition $\omega_e\tau_{2e}=1$.}
\label{fg:pre}
\end{figure*}

The terms in the last brackets in equations~(\ref{eq:T1PRE}) through (\ref{eq:T2PRE_c}) are the spectral density functions $J(\omega)$ that provide a measure of the likelihood of the random fluctuations at a frequency $\omega$.\cite{pr_73_679_1948} The nuclear longitudinal relaxation rate is proportional to the spectral density function at the nuclear Larmor frequency, while the transverse relaxation rate has an additional term which includes the spectral density at zero frequency. The dependence of the nuclear relaxation times on the correlation time is shown in Fig.~(\ref{fg:pre}). In the motional narrowing regime\cite{pomr} $T_1$ passes through a minimum for  $\omega_n\tau_{1e}=1$, while $T_2$ decreases monotonically with increasing correlation times, i.e., with slower motions.

In addition, fluctuations of the Fermi contact interaction can also be a source of relaxation, which could be relevant for nuclei in the vicinity of the paramagnetic metal ions. Generally, care should be taken when treating nuclei in the first coordination shells, due to the very large coupling strengths, $\delta$, and slow electron relaxation times, it is likely that the condition of motional narrowing ($\delta\tau_c\ll1$) is not fulfilled.\cite{pomr} In any case, in samples with dilute paramagnetic centres, as in MI-DNP applications, the fraction of nuclei experiencing significant FC interactions is small and we will assume their contribution to the macroscopic relaxation behaviour negligible.

\begin{figure*}
\begin{center}
\includegraphics[scale=1]{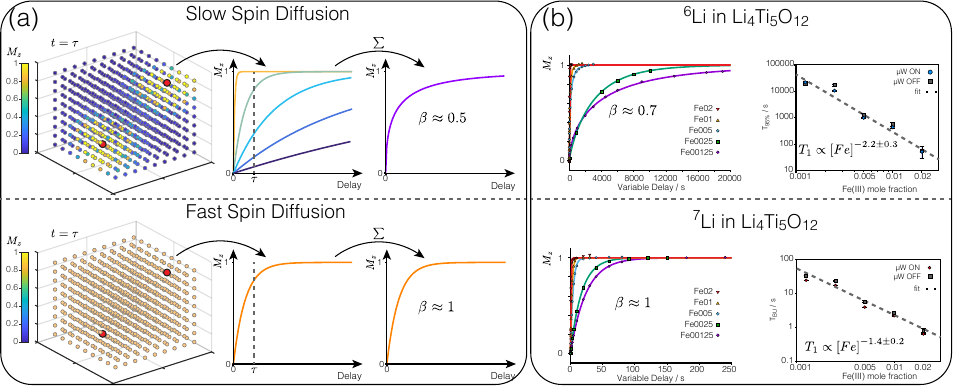}
\end{center}	
\caption{{\bf Paramagnetic relaxation enhancement with and without spin diffusion.} (a) Top, in the absence of efficient spin diffusion and intrinsic relaxation mechanisms, the presence of paramagnetic dopants, will lead to a distribution of relaxation times, following the distribution of nucleus-electron distances (in the figure the angular dependence is neglected). In a saturation recovery $T_1$ measurement, the added contribution of many single exponential recovery curves results approximately in a stretched exponential build-up,  see equation \ref{eq:stretchedT1}, with the stretched exponent $\beta$ approaching 0.5 in a 3D sample. Bottom, in the limit of fast spin diffusion, polarization transfer among nuclear spins is faster than the relaxation process, thus all spins relax with a common relaxation time and the added contributions follow a single exponential curve.
(b) Experimental magnetization recovery curves (left) and longitudinal relaxation times vs. concentration (right) for $^6$Li (top) and $^7$Li (bottom) in Fe(III) doped Li$_4$Ti$_5$O$_{12}$. Spin diffusion is weaker in $^6$Li due to its lower gyromagnetic ratio and natural abundance, leading to a markedly distinct relaxation behaviour of both nuclei. The fit parameter are in good agreement with the expected behaviour in slow and fast spin diffusion regimes, respectively. Data in (b) adapted with permission from data given in References~\cite{jpcc_124_7082_2020,jpcl_11_5439_2020}. }
\label{fg:spindiff}
\end{figure*}

 These equations describe the relaxation rate for each nucleus due to the coupling to a single electron spin. Due to the strong distance dependence ($r^{-6}$ for  relaxation via through-space dipolar couplings), in a sample with dilute paramagnetic centres it is a good approximation to assume that the relaxation rate of the nuclear spins is dominated by isolated paramagnetic centres. At the same time, each paramagnetic centre icontributes to the relaxation of many nuclei. Consequently, nuclear relaxation times will differ strongly throughout the sample, due to the large variation in $\omega_d^{en}$. This is shown schematically in Fig.~(\ref{fg:spindiff}a, top). While nuclei experiencing the same coupling strength will relax to equilibrium following an exponential curve, as predicted by the Bloch equations (see Appendix), the added contribution of all nuclear spins $N_n$ can be approximated by a stretched exponential function.\cite{prb_74_184430_2006} In a saturation recovery experiment, this leads to:
 \begin{equation}
M_z= 
\sum_i^{N_n} [1-\exp[-(tR_{1,i})]]
\approx
M_{eq}[1-\exp[-(tR_{1})^\beta]].
\label{eq:stretchedT1}
 \end{equation}
 where the contribution of each spin is normalized and $\beta$ is the stretched or Kohlrausch exponent and approaches 0.5 in a 3D sample with homogeneously distributed paramagnetic centres.\cite{prl_21_511_1968} Analogously, the transverse relaxation can be described by a stretched exponential decay. Further, the relaxation rates depend on the metal ion concentration $[M]$ as $R_{1,2}\propto[M]^2$.\cite{prl_65_614_1990,prb_54_15291_1996}

 So far, in the discussion of the paramagnetic relaxation enhancement, we have only considered the case of isolated nuclear spins, each relaxed directly by a paramagnetic centre. In a network of strongly coupled homonuclear spins, the relaxation behaviour of the sample will differ significantly. In this case, polarization can be transferred between nuclei, mediated by the flip-flop term $\left(\hat{I}_+\hat{I}_-+\hat{I}_-\hat{I}_+
\right)$  of the homonuclear dipolar coupling. In a macroscopic ensemble this effect tends towards homogenizing the polarization across the sample and is termed spin diffusion (we will treat the concept of spin diffusion in more depth in section~\ref{subsec:polvsRelax}). In the presence of fast spin diffusion, the homogenization of the polarization occurs at a faster rate compared to the relaxation rates described in equations~(\ref{eq:T1PRE}) and (\ref{eq:T2PRE}). Consequently, homogenization of the polarization occurs at a much faster rate than the direct relaxation processes and the macroscopic polarization relaxes following a single exponential function,\cite{p_15_386_1949} as depicted in Fig.~(\ref{fg:spindiff}a, bottom). In this regime the relaxation rate is directly proportional to the concentration of dopants:  $R_{1,2}\propto[M]$.\cite{pr_119_79_1960,pr_166_292_1968} A more complex behaviour is expected in intermediate regimes, where direct relaxation governs the behaviour of nuclei close to the paramagnetic centre, while spin diffusion is more relevant for remote nuclei.\cite{pr_119_79_1960,pr_166_279_1968} It is also possible for longitudinal and transverse relaxation to be in different regimes, if their rates are very different.\cite{jmr_336_107143_2022}

In Fig.~(\ref{fg:spindiff}b) the longitudinal relaxation behaviours of $^6$Li and $^7$Li in Li$_4$Ti$_5$O$_{12}$ obtained experimentally are shown.\cite{jpcc_124_7082_2020,jpcl_11_5439_2020} Clear differences in both, stretched factor $\beta$ and concentration dependence were observed, as expected due to the large ratio of the gyromagnetic moments of  $^6$Li and $^7$Li ($\sim$factor~2.6) and of the isotopic abundance ($\sim$factor~12), as well as quadrupolar coupling strength, leading to a stronger coupling network and thus more efficient spin diffusion for the latter. Another consequence of the presence of efficient spin diffusion is that the relaxation times are significantly shorter (i.e. shorter than would be expected only from the ratio of the gyromagnetic ratios).\\

An interesting property of equations~(\ref{eq:T1PRE_c}) and (\ref{eq:T2PRE_c}) that becomes evident when looking at Fig.~(\ref{fg:pre}) is that the ratio of longitudinal and transverse relaxation times ($T_1$/$T_2$) is independent of the coupling strength between the nucleus and electron. Instead, the ratio depends only on the nuclear Larmor frequency and the correlation time, $\tau_{1e}$, according to:\cite{jmr_336_107143_2022}
\begin{equation}
\frac{T_1}{T_2}=\frac{7}{6}+\frac{4}{6}\tau_{1e}^2\omega_n^2.
\label{eq:T1T2ratio}
\end{equation}
This, of course, will only be valid as long as both, $T_1$ and $T_2$, are dominated by the same relaxation mechanism and both are in the same spin diffusion regime. However, under these conditions, one can easily obtain the size of $\tau_{1e}$:
\begin{equation}
 \tau_{1e}=\sqrt{\left(\frac{T_1}{T_2}-\frac{7}{6}\right)\frac{6}{4\omega_n^2}}.
\label{eq:T1efromT1T2}
\end{equation}
It is important to emphasize that this equation is not valid in the extreme narrowing regime of relaxation, where the ratio of $T_1$/$T_2$ is constant and independent of $\tau_{1e}$ (see Fig.~(\ref{fg:pre})). At least at low dopant concentrations, when electron-electron interactions are weak, $\tau_{1e}$ is likely a good estimate of the electron relaxation time $T_{1e}$. This way, equation~(\ref{eq:T1efromT1T2}) enables one to estimate $T_{1e}$ experimentally in a very simple way. As we will see later, $T_{1e}$ is a fundamental parameter for determining the success of a DNP experiment, and it can be very difficult to assess by other experimental techniques, particularly at conditions of interest for MI-DNP, high magnetic fields and temperatures.

\begin{figure*}
\begin{center}
\includegraphics[scale=1]{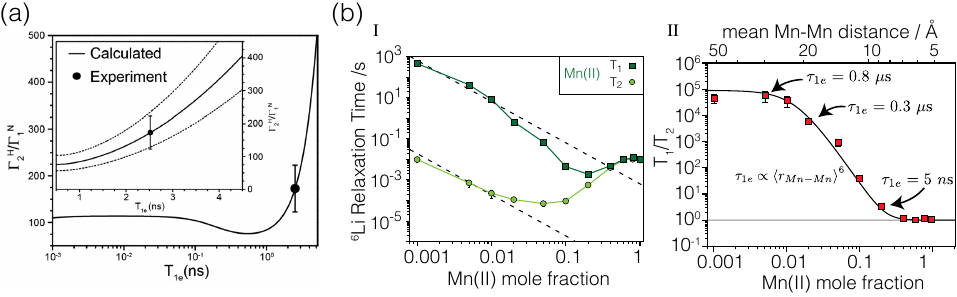}
\end{center}	
\caption{{\bf Obtaining $\tau_{1e}$ from the ratio of nuclear ${\bf T_1}$/${\bf T_2}$.} (a) Calculated ratio of $R_2$ of $^1$H over $R_1$ of $^{15}$N  coupled to the same paramagnetic Cu$^{2+}$ ion, either at the same distance (solid line) or at a distance variation of $\pm$5\% (dashed lines) and as a function of $T_{1e}$ (in figure symbol for the rates is $\Gamma_{1,2}$). The line passes a minimum due to the different Larmor frequencies of both nuclei. The solid line was used to estimate Cu$^{2+}$ $T_{1e}$ from the experimental relaxation rates ratio (black dot) of backbone amide in protein GB1 K28C-EDTA-Cu$^{2+}$ mutant. (b) Left: measured $^6$Li $T_1$ and $T_2$ relaxation times in LiMg$_{1-x}$Mn$_x$PO$_4$ for varying Mn(II) content. Right: Corresponding ratio of $T_1$/$T_2$ with calculated  $\tau_{1e}$ for some selected points, using equation~\ref{eq:T1efromT1T2}. Reproduced (a) and adapted (b) with permissions from References~\cite{jpcl_8_5871_2017} and \cite{jmr_336_107143_2022}, respectively.}
\label{fg:T1eT1T2}
\end{figure*}

To our knowledge, the first use of this simple relation for determining $T_{1e}$ was published by Mukhopadhyay et al.\cite{jpcl_8_5871_2017} and is shown in Fig.~(\ref{fg:T1eT1T2}a). The shape of the calculated curve does not follow equation~(\ref{eq:T1T2ratio}), because the $R_1$ and $R_2$ rates were obtained for different nuclei, $^1$H and $^{15}$N, respectively. A systematic study of the nuclear longitudinal and transverse relaxation times as a function of the concentration of metal ions was carried out by the author's group and is shown in Fig.~(\ref{fg:T1eT1T2}b).\cite{jmr_336_107143_2022} By mapping the ratio of $T_1$/$T_2$ of $^6$Li in LiMg$_{1-x}$Mn$_x$PO$_4$ it was possible to observe changes in $\tau_{1e}$ due to increasingly stronger interactions between paramagnetic Mn(II) centres. At low concentrations ($x<0.01$) both $T_1$ and $T_2$ decrease with the expected $[M]^{-2}$ dependence, thus their ratio stays constant. Above this sample- and metal-specific threshold of dopant content, electron-electron interactions enhance the rate of fluctuations of the electron spins,leading to a large drop in the ratio of the nuclear relaxation times $T_1$/$T_2$, a deviation from the $T_{1,2}\propto[M]^{-2}$ relation, and at some point even to an increase of the relaxation times with increasing paramagnetic dopant concentration. While the nature of the decrease in $\tau_{1e}$ is not  trivial and could be due to either coherent or non-coherent effects, the ratio of $T_1$/$T_2$ decreases with the mean Mn-Mn distance as $\tau_{1e}\propto\left \langle r_{Mn-Mn}\right \rangle^{6}$, which could be indicative that in this case the fluctuations of the electron magnetic moment are caused by relaxation mediated by through-space dipolar couplings between the electron spins.

    \subsection{Magnetic susceptibility}
            \label{subsec:susceptibility}

\begin{figure*}
\begin{center}
\includegraphics[scale=1]{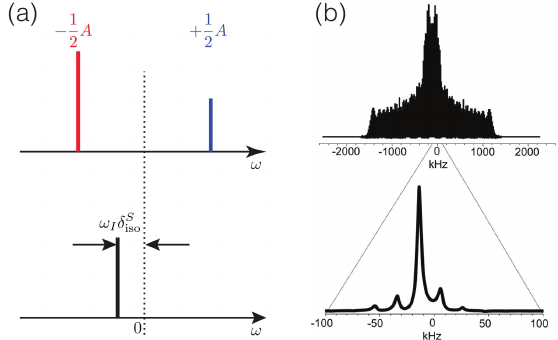}
\end{center}	
\caption{{\bf In the limit of fast $T_{1e}$, nuclei sense an averaged magnetic moment.} (a) Schematic representation of an NMR spectrum of a nucleus coupled to an electron, both spin 1/2, with a hyperfine coupling $A_{FC}$, in the absence of electron relaxation (top) and in the limit of very fast relaxation (bottom). The difference in the intensity of both peaks in the doublet arise from a difference in the population, due to the difference in Zeeman energy of the electron spin. The averaged resonance is shifted to the centre of mass of the doublet. (b) Simulated MAS spectrum of a proton nucleus coupled to an electron spin 1/2 via isotropic FC and anisotropic through space dipolar coupling, without (top) and with fast (bottom) electron spin relaxation. Averaging of the FC leads to an isotropic shift, while averaging of the dipolar coupling results in a significantly narrower inhomogeneous powder pattern. Reproduced with permission from Ref.~\cite{jmr_326_106939_2021}.}
\label{fg:pNMR_theory}
\end{figure*}

In the limit of rapid electron relaxation, the dipolar (or contact) coupling does not lead to any multiplet splitting of the nuclear resonance frequency, but rather to a shift (see Fig.~\ref{fg:pNMR_theory}).\cite{pnmrs_111_1_2018} The magnitude of this paramagnetic frequency shift is given by the  average frequency of the multiple transitions, weighted by their relative contributions. The populations of the multiplet components differ mainly due to the difference in the electron Zeeman energy. In other words, this can be understood as if the nuclear spins were experiencing an additional local magnetic field, the size of which is given by the average magnetic moment of the electron spins, which in turn can be related to the magnetic susceptibility. In complete analogy to exchange narrowing,\cite{sdbonmr} the validity of the rapid relaxation limit can be evaluated from the ratio of the electron relaxation rate (equivalent to the exchange rate) and the magnitude of the coupling (equivalent to the frequency difference).

It is often useful to describe the effects of the paramagnetic centres on the NMR spectrum in terms of the magnetic susceptibility. In the following we will briefly introduce the relevant concepts. Placing a material in a magnetic field, ${\bf H}$, induces a magnetization ${\bf M}$, and the total magnetic induction, ${\bf B}$, becomes:
      \begin{equation}
{\bf B}=\mu_0({\bf H}+{\bf M}),
  \end{equation}
where $\mu_0$ is the vacuum permeability constant. The sign and magnitude of the induced magnetization depends on the size of the external magnetic field ${\bf H}$ and on a property of the material called the magnetic susceptibility. The magnetic susceptibility per unit volume, $\chi_v$, is unitless and defined according to:
      \begin{equation}
     {\bf M}= \chi_v{\bf H},
  \end{equation}
and takes negative values for diamagnetic materials, and positive for paramagnetic and ferro- and antiferromagnetic materials. In this text we will focus only on paramagnetic systems, where  $\chi_v\ll1$,\cite{pnmrs_40_249_2002} thus ${\bf B}\approx\mu_0{\bf H}$ and throughout the review when using the term `magnetic field' we will be referring to ${\bf B}$.\cite{pnmrs_111_1_2018} The molar susceptibility is obtained from the susceptibility per unit volume by multiplying with the molar volume: $\chi_m=V_m\chi_v$. The origin of the paramagnetic susceptibility is associated with the magnetic moments of the electron spin and angular momentum (neglecting  contributions from the much smaller nuclear paramagnetism). For the simplest case, the longitudinal component of the magnetic moment $\hat{\mu}_z$ can take values from $-m_S\hbar\gamma_e$ to $m_S\hbar\gamma_e$. Ultimately, we wish to find the relation between the susceptibility, which is a macroscopic property, with the average magnetic moment, $\langle\hat{\mu}_z\rangle$, of each electron spin, a microscopic property. The magnetic moment is related to the spin according to:
 \begin{equation}
{\boldsymbol{ \hat{\mu}}}=-\mu_B g_e \bf{\hat{S}}.
  \end{equation}
The mean magnetic value of an electronic spin is obtained from the expectation value of the longitudinal component of the spin $\langle\mu_z\rangle$.\cite{pnmrs_40_249_2002} One can evaluate the expectation value between two levels of known energy according to Boltzmann statistics. In the high temperature approximation $\left(\frac{\Delta E}{kT}\ll1\right)$ we obtain:
 \begin{equation}
\langle\hat{S}_z\rangle=-\frac{\mu_Bg_eB_0S(S+1)}{3kT},
  \end{equation}
and
\begin{equation}
\langle\hat{\mu}_z\rangle=\frac{\mu_B^2g_e^2B_0S(S+1)}{3kT}.
\end{equation}
The magnetization can be obtained from the sum over the mean magnetic moments of all electron spins:
\begin{equation}
M=\frac{N_A}{V_m}\langle\hat{\mu}_z\rangle=\frac{N_A}{V_m}\frac{g_e^2\mu_B^2B_0S(S+1)}{3kT}.
\end{equation}
With the previous equations we finally obtain:
\begin{equation}
\chi_m=\frac{\mu_0N_A\langle\hat{\mu}_z\rangle}{B_0}
=\frac{\mu_0\mu_B^2g_e^2N_AS(S+1)}{3kT}.
\end{equation}

The high temperature approximation ensures that the magnetization is  proportional to the external magnetic field. This condition might be violated at very low temperatures or very high fields. In addition, we have assumed so far an ideal system of isolated spins following the Curie law. Under these assumptions, the magnetic susceptibility increases linearly with the inverse of the temperature and becomes zero as the temperature approaches infinity. With increasing concentration of paramagnetic centres interactions between spins become important and the system will deviate from a purely paramagnetic behaviour. Strong exchange coupling between electron spins can lead to a macroscopic ordering of the magnetic moments of the electron spins throughout the sample even in the absence of an external magnetic field. Ordering into a ferromagnetic or antiferromagnetic state occurs when cooling below a transition temperature, named Currie or N\'{e}el temperatures, respectively. Consequently, the zero crossing of the magnetic susceptibility will be shifted. This is taken into account by the Weiss constant $\Theta$ in the Curie-Weiss law, and we can rewrite the expressions of the expectation value of S$_z$ and of the molar susceptibility according to:\cite{pnmrs_111_1_2018}
 \begin{equation}
\langle\hat{S}_z\rangle=-\frac{\mu_Bg_eB_0S(S+1)}{3k\left(T-\Theta\right)},
  \end{equation}
  and
\begin{equation}
\chi_m
=\frac{\mu_0\mu_B^2g_e^2N_AS(S+1)}{3k\left(T-\Theta\right)}.
\end{equation}

Now we can turn our attention back to the coupling between the electron spins and the nuclear spins. As previously mentioned, the consequence of this averaging from the point of view of a nucleus is that it will sense an average magnetic moment that has the same form as the shielding interaction. Consequently, the isotropic Fermi contact shift will manifest as an isotropic chemical shift,\cite{jcp_29_1361_1958} while the through-space dipolar coupling will result in an interaction that can be described by a tensor equivalent to the chemical shift anisotropy.\cite{jcp_89_4600_1988} Further, it is possible to relate the shifts to the macroscopic magnetic susceptibility, since both depend on the average value of the magnetic moment of individual spins. The expressions for the paramagnetic isotropic and anisotropic shielding are given by:\cite{pnmrs_111_1_2018}
\begin{equation}
\sigma_{iso}
=\frac{\langle\hat{S}_z\rangle}{\gamma_IB_0}A_{FC}
=\frac{\chi_m}{\gamma_I\mu_0\mu_Bg_eN_A}A_{FC}
=-\frac{\mu_Bg_eS(S+1)}{3\gamma_Ik\left(T-\Theta\right)}A_{FC}.
\end{equation}
and
\begin{equation}
\sigma_{aniso}(\theta,\phi)
=\frac{\langle\hat{S}_z\rangle}{\gamma_IB_0}\omega_d^{en}{\bf D}(\theta,\phi)
=\frac{\chi_m}{\gamma_I\mu_0\mu_Bg_eN_A}
\omega_d^{en}{\bf D}(\theta,\phi)
=-\frac{\mu_Bg_eS(S+1)}{3\gamma_Ik\left(T-\Theta\right)}\omega_d^{en}{\bf D}(\theta,\phi)
.
\end{equation}
Here ${\bf D}(\theta,\phi)$ is again the tensor of the dipolar coupling (equation \ref{eq:Dtensor}).\cite{jpcb_106_3576_2002} The paramagnetic NMR shifts in Hz can be obtained from:
\begin{equation}
\Delta\nu_{iso}
=\frac{\gamma_I\sigma_{iso}B_0}{2\pi}
=-\frac{\mu_Bg_eB_0S(S+1)}{6\pi k\left(T-\Theta\right)}A_{FC}.
\end{equation}
and
\begin{equation}
\Delta\nu_{aniso}(\theta,\phi)
=\frac{\gamma_I\sigma_{aniso}B_0}{2\pi}
=-\frac{\mu_Bg_eB_0S(S+1)}{6\pi k\left(T-\Theta\right)}\omega_d^{en}{\bf D}(\theta,\phi).
\end{equation}

From these equations one can see that the shifts scale linearly with the magnetic field (in analogy to chemical shifts) and with the inverse of the temperature, following the behaviour of the mean magnetic moment in the high temperature approximation.

We should emphasize that in the section about PRE we have taken into account the full magnetic moment of the electron spin rather than its averaged value. The fluctuation of the full magnetic moment due to the electron relaxation is actually the source of relaxation. However, an additional PRE mechanism can arise from fluctuations of the interaction between the nuclei and the averaged electron magnetic moment and is called Curie relaxation.\cite{jmr_19_58_1975} However, it is unlikely that correlation times from motions in solids will be sufficiently fast to make this relaxation mechanism significant, therefore will not be considered further in this text.

 Finally, we note that the presence of paramagnetic centres can also lead to shifts and shift anisotropies due to bulk magnetic susceptibility and anisotropic bulk magnetic susceptibility effects.\cite{pr_79_179_1950,pr_81_717_1951,jmr_119_157_1996,jmr_133_330_1998,jmr_234_44_2013,jacs_141_13089_2019} While the former effect can be removed by MAS, the latter can lead to line broadening in polycrystalline samples, even under MAS. Although in the presence of slow-relaxing electrons, this effect is likely masked by short $T_2$ relaxation times from PRE.

    \subsection{The NMR spectrum}
            \label{subsec:pNMRexamples}

\begin{figure*}
\begin{center}
\includegraphics[scale=1]{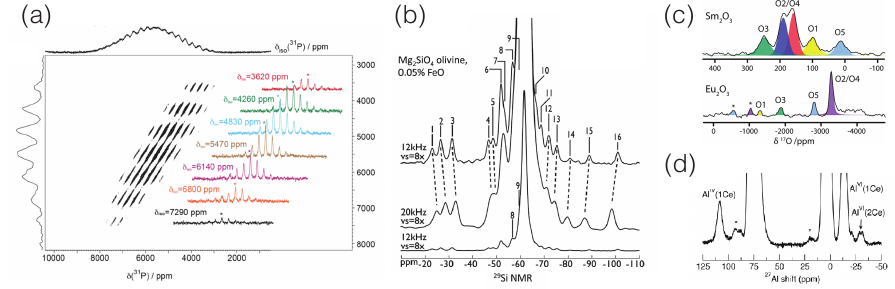}
\end{center}	
\caption{{\bf NMR spectra in the presence of paramagnetic centres.} (a)~$^{31}$P aMAT 2D spectrum of LiFe$_{0.5}$Mn$_{0.5}$PO$_{4}$. Vertical axis is the purely isotropic dimension and horizontal axis the MAS dimension; projections of each are shown alongside the axes. Data obtained at 320~K, 11.7~T and a MAS speed of 60~kHz. Reproduced with permission from Ref.~\cite{jacs_134_17178_2012}. (b) $^{29}$Si spectra of Mg$_2$SiO$_4$ doped with 0.05\% FeO. Data obtained at 14.1~T and a MAS speed of 12 and 20~kHz, for the latter heating from air friction due to spinning was estimated to raise temperature up to 50$^\circ$C. Reproduced with permission from Ref.~\cite{ac_73_128_2017}, (c) $^{17}$O spectra of: Top: Sm$_2$O$_3$ at -44$^\circ$C, 9.4T and MAS speed of 30~kHz. Bottom: Eu$_2$O$_3$ at room temperature, 4.7~T and MAS speed of 60~kHz. Reproduced with permission from Ref.~\cite{ssnmr_102_21_2019}. (d) $^{27}$Al spectrum of Y$_{2.91}$Ce$_{0.09}$Al$_5$O$_{12}$ acquired at room temperature, 23.5~T and a MAS speed of 60~kHz. Reproduced with permission from Ref.~\cite{cm_25_3979_2013}.}
\label{fg:pNMR}
\end{figure*}

\begin{figure*}
\begin{center}
\includegraphics[scale=1]{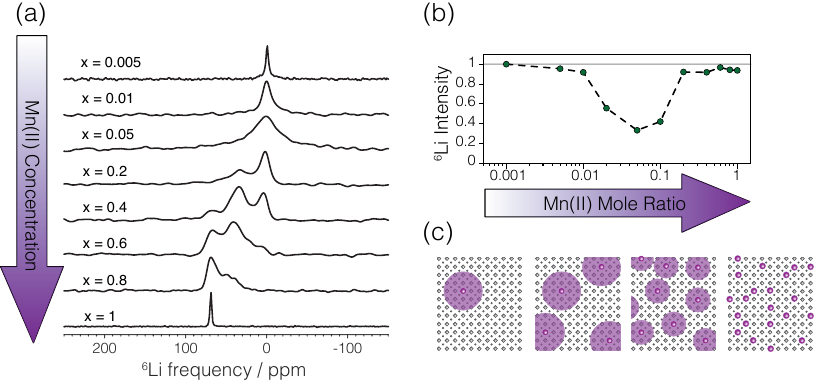}
\end{center}	
\caption{{\bf Effect of concentration of paramagnetic centres on line broadening and signal quenching.} (a)~$^6$Li NMR Hahn echo spectra normalised according to the maximum intensity  of LiMg$_{1-x}$Mn$_x$PO$_4$ for varying Mn(II) content. (b) Quantification of the corresponding signal intensity, deviation from unity is due to paramagnetic signal quenching. (c) Schematic representation of signal quenching with increasing Mn(II) content. At low concentrations, the quenching sphere (purple circle) around each individual Mn(II) ion is large. With increasing concentration, while the number of paramagnetic centres grows, at some point the size of each sphere shrinks, eventually enabling acquisition of nuclear spins in the immediate proximity of the paramagnetic centres. Combination of these two opposing effects leads to the minimum of $^6$Li signal intensity in (b). Adapted with permission from Ref.~\cite{jmr_336_107143_2022}.}
\label{fg:quench}
\end{figure*}

    The NMR spectrum of a sample with paramagnetic centres will depend strongly on the correlation time of the fluctuations of the electron magnetic moment $\tau_{1e}$,\cite{jmr_336_107143_2022} which is often equal to $T_{1e}$ (see preceding discussion). For fast fluctuations ($\tau_{1e}<10^{-9}\text{ s}$) NMR relaxation will be in the extreme narrowing regime,\cite{ponm} where $T_1\approx$~$T_2$, as evident from Fig.~(\ref{fg:pre}). In this regime the relatively mild $T_2$ line broadening enables reasonable resolution. When the paramagnetic shifts are more pronounced than the broadening, valuable chemical information can be obtained from simple spectra.  With increasingly long $\tau_{1e}$ times, the nuclear $T_2$ relaxation times will decrease, leading to broader lines, lower resolution and eventually signal quenching. In the following we will discuss the type of  information that can be gained from NMR spectra in the presence of paramagnetic centres and how this can be useful for evaluating candidates for MI-DNP. To this end we will look at a few examples, focusing on inorganic oxides. These represent only a few selected studies within the very large area of paramagnetic NMR in materials science. In Fig.~(\ref{fg:pNMR}) some examples of cases in the short $\tau_{1e}$  regime are shown, while in Fig.~(\ref{fg:quench}) the changes observed in the NMR spectrum due to changes of $\tau_{1e}$ from fast to slow regime are illustrated.
    
    In Fig.~(\ref{fg:pNMR}) NMR spectra of samples with paramagnetic metal ions from the d-block (Fig.~(\ref{fg:pNMR} a and b)) and from the f-block (Fig.~(\ref{fg:pNMR} c and d)) are shown. In Fig.~(\ref{fg:pNMR} a and c), the paramagnetic centres are a major component of the composition, thus the samples are highly paramagnetic, while in (b) and (d) they are introduced as minor dopants. In all of these cases, short $\tau_{1e}$ guarantees relatively long $T_2$, and nuclear spins that are only one or two bonds away from the paramagnetic metal centres can be detected. The $^{31}$P MAS spectrum of LiFe$_{0.5}$Mn$_{0.5}$PO$_{4}$ shows a very broad peak due to both a distribution of isotropic paramagnetic Fermi contact shifts, and the presence of large anisotropic paramagnetic dipolar broadening.\cite{jacs_134_17178_2012} MAS at 60~kHz is not sufficient to fully remove the anisotropic contributions, leading to many overlapping sidebands. Cl{\'{e}}ment et al. acquired a series of 2D adiabatic magic angle turning (aMAT) spectra for various ratios of Fe/Mn in the structure (Fig.~(\ref{fg:pNMR} a)), allowing the isotropic resonance to be distinguished from the MAS side-band manifold. From these spectra they were able to obtain detailed information on the electronic structure around the phosphorous atoms and discern all possible pathways connecting them to the metal centre. For Sm$_2$O$_3$ and Eu$_2$O$_3$ $^{17}$O spectra were measured at various temperatures to gain insights into the temperature dependence of the magnetic susceptibility (Fig.~(\ref{fg:pNMR} c)).\cite{ssnmr_102_21_2019} Remarkably, Figs.~(\ref{fg:pNMR} b and d) show how paramagnetic NMR is capable of characterizing structural changes surrounding doping sites at concentrations as low as 0.05\% (here, FeO in Mg$_2$SiO$_4$ and Ce(III) in Y$_{3}$Al$_5$O$_{12}$). Nuclei in close proximity to the paramagnetic centres are shifted due to Fermi contact and pseudocontact shifts enabling the assignment of site occupancy.\cite{ac_73_128_2017,cm_25_3979_2013}

Fig.~(\ref{fg:quench} a) shows the $^6$Li Hahn spin-echo NMR spectra of LiMg$_{1-x}$Mn$_x$PO$_4$, for $x=0.005$ to $x=1$.\cite{jmr_336_107143_2022} We will base the discussion on the results introduced in section \ref{subsec:pre}, where we had seen that $\tau_{1e}$ of Mn(II) decreases with increasing dopant concentration in this sample, from ca. 1~$\mu$s at very low concentrations $x$ to less than 1~ns, for values of $x>0.1$. Upon increasing manganese concentration, manganese will substitute for magnesium in the lattice without altering the crystallographic structure. At the lowest Mn(II) content a single sharp peak is observed, which becomes broader with increasing dopant concentration up to $x=0.05$. Above this concentration, the signal becomes sharper again and multiple peaks appear, owing to a superpositions of different FC couplings.\cite{jacs_134_17178_2012} At $x=1$ again only a single sharp signal is observed. In Fig.~(\ref{fg:quench} b) the $^6$Li integrated intensity is shown, which follows the same trend as the line broadening, where at $x=0.05$ a minimum of the integrated intensity as low as 30\% is observed. This behaviour can be rationalize by following the changes in $\tau_{1e}$. Long $\tau_{1e}$ values result in very short nuclear $T_2$ times (equation~(\ref{eq:T2PRE_c})), to the point that the coherence lifetime of nuclear spins in close proximity to Mn(II) is shorter than the echo delay. The signal that cannot be detected due to its fast decay is said to be quenched, which is represented in Fig.~(\ref{fg:quench} c) by the purple circles surrounding the paramagnetic centres. At first, with increasing Mn(II) content the total volume containing quenched resonances increases. However, at some point as  $\tau_{1e}$ decreases, so taht the quenching radius of each Mn(II) ion shrinks, because the $T_2$ of nuclei at a fixed distance from the paramagnetic centre becomes larger. Ultimately $\tau_{1e}$ becomes so short and the resulting quenching radius so small, that all lithium nuclei in the sample can be detected.
In MI-DNP long $\tau_{1e}$ values are required, therefore, signal quenching can be expected and it is important to be aware of this when comparing absolute sensitivities for an NMR experiment (see appendix).

\section{The DNP experiment}
         \label{sec:dnp}
Since the properties of the metal ion dopants will depend on the properties of the lattice, every sample has different DNP properties for MI-DNP. It is therefore useful to get familiarized with the basic theoretical concepts of DNP as this will help understand possible differences in the DNP behaviour of each sample and assist in the quest for optimal conditions. Here we will follow the formalism and nomenclature introduced by Hovav et al.\cite{jmr_207_176_2010} For simplicity, we will consider only spins 1/2. We note, however, that in most examples of MI-DNP paramagnetic metal ions with spin higher than 1/2 were used. We explain this by considering the fact that DNP occurs mainly due to the central transition of the electron spins. In the next section we will briefly mention some of the consequences. For a more thorough discussion of the use of high-spin metals for DNP we refer to a review by Corzilius.\cite{emagres_7_179_2018}

We start our discussion by introducing matrix representations of the relevant Hamiltonians for a coupled two-spin system consisting of one nucleus and one electron to justify the high-field approximation. Next, we consider the solid effect (SE) DNP mechanism by including the effect of microwave irradiation. Besides the solid effect, the main other mechanisms at high magnetic fields are the Overhauser effect (OE) and the cross effect (CE) mechanisms. As to date these seem to be less relevant for DNP using metal ions, we will only briefly introduce the main concepts and refer to Refs.~\cite{amr_3_79_1968,pnmrs_74_33_2013,jmr_264_78_2016} and \cite{jmr_214_29_2012,jcp_137_084508_2012,jmr_224_13_2012,jmr_258_102_2015,emr_8_295_2019}, respectively, for a deeper theoretical treatment.

    \subsection{The Spin Hamiltonian in a DNP experiment}
        \subsubsection{A coupled electron-nuclear two spins system}
The description of the solid effect DNP mechanism requires only two spins, one nucleus and one electron, which are coupled through the dipolar interaction. The spin Hamiltonian of this simple system only includes the Zeeman interactions of the electron spin S and nuclear spin I, which depend on their respective gyromagnetic ratios and the external magnetic field, as well as the dipolar coupling Hamiltonian, described in equation~\ref{eq:dipalph}:
\begin{equation}
\hat{H}_{0}=\hat{H}_{eZ}+\hat{H}_{nZ}+\hat{H}_{dd}^{en},
\label{eq:Hspin}
\end{equation}
with
\begin{equation}
\begin{split}
\hat{H}_{eZ}=&-\gamma_e B_0\hat{S}_z=\omega_e\hat{S}_z,\\
\hat{H}_{nZ}=&-\gamma_n B_0\hat{I}_z=\omega_n\hat{I}_z.
\end{split}
\end{equation}

\begin{figure*}
\begin{center}
\includegraphics[scale=1]{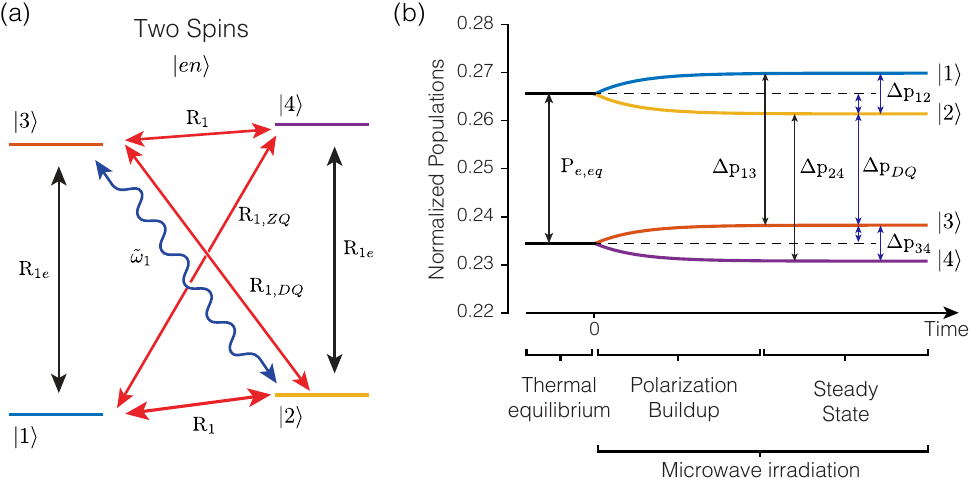}
\end{center}	
\caption{{\bf Irradiation of the DQ transition in solid effect DNP.} (a) Energy level diagram of  of a two spin $e-n$ system. Microwave irradiation on the DQ transition is shown as wavy arrow and relaxation paths are indicated by straight arrows. (b) Effect of irradiating the DQ transition on the populations. At thermal equilibrium the difference in population due to the nuclear Zeeman interaction is negligible in the shown scale. When irradiating on the DQ transition starting at $t=0$, populations of states  \bra{}{3} and \bra{}{2} become closer (at sufficient high power they equalize, and the transition is said to be fully saturated). Due to the efficient electron relaxation, the ratios p$_1$/p$_3$=p$_4$/p$_2$ remain constant, leading to a decrease in the population of \bra{}{4} and increase in the population of \bra{}{1}. }
\label{fg:enhpop}
\end{figure*}

Fig.~(\ref{fg:enhpop}~a) shows schematically the four possible energy levels, which in the pure Zeeman eigenbasis are:
\begin{equation}
\begin{split}
\bra{}{1}&=\bra{}{\beta\alpha} \qquad
\bra{}{2}=\bra{}{\beta\beta}\\
\bra{}{3}&=\bra{}{\alpha\alpha}\qquad
\bra{}{4}=\bra{}{\alpha\beta}.
\end{split}
\end{equation}
The first element refers to the electron spin and the second to the nuclear spin. In the Zeeman basis the matrix representation of the spin Hamiltonian is:
\begin{equation}
\hat{H}_{0}=\frac{1}{2}
\begin{pNiceMatrix}[first-row,last-col=5]
 \bra{}{\alpha\alpha}   & \bra{}{\alpha\beta}   &\bra{}{\beta\alpha}    &\bra{}{\beta\beta} &\\
 \omega_e+\omega_n +\frac{1}{2}A\omega_{d}^{en} & C\omega_{d}^{en} &  C\omega_{d}^{en} &2E\omega_{d}^{en} &\ket{}{\alpha\alpha}\\
 D\omega_{d}^{en}  & \omega_e-\omega_n-\frac{1}{2}A\omega_{d}^{en}  &2B\omega_{d}^{en} & -C\omega_{d}^{en}&\ket{}{\alpha\beta}\\
   D\omega_{d}^{en}  & 2B\omega_{d}^{en} &-\omega_e+\omega_n-\frac{1}{2}A\omega_{d}^{en} &- C\omega_{d}^{en} &\ket{}{\beta\alpha}\\
  2F\omega_{d}^{en}  & - D\omega_{d}^{en} &-D\omega_{d}^{en}& -\omega_e-\omega_n+\frac{1}{2}A\omega_{d}^{en} &\ket{}{\beta\beta}
\end{pNiceMatrix}
.
\end{equation}
Note that the gyromagnetic ratio of the electron is negative, while for most nuclei it is positive. Thus, for a positive $\gamma_n$, $\omega_n$ is negative and the state $\bra{}{\beta\alpha}$ is the lowest in energy.

    \subsubsection{Microwave irradiation}
The spin system can be manipulated by the application of external oscillating fields in the radio-frequency (rf) and microwave ($\mu$w) range, affecting nuclear and electron spins, respectively. Describing the SE DNP mechanism only requires accounting for the effect of microwave irradiation perpendicular to the external magnetic field (see Fig.~\ref{fg:enhpop}). At this point it is convenient to go into the rotating frame representation, in which the $\mu$w irradiation appears static and will be defined to be along the x-direction:
\begin{equation}
\tilde{\hat{H}}_{\mu w}=\omega_1\hat{S}_x,
\label{eq:MW}
\end{equation}
the tilde on the Hamiltonian indicates that it is in the rotating frame representation.

        \subsubsection{The pseudo-secular approximation}
In NMR it is often sufficient, and convenient, to treat the spin system within the secular approximation. In the secular approximation all terms which do not commute with the Zeeman interaction are truncated. 
In the case of the electron-nuclear spin system the secular approximation might not be a good approximation, as the dipolar coupling can be of the same order of magnitude as the nuclear Zeeman interaction. Instead, only terms that do not commute with the electron Zeeman interaction will be truncated, a procedure known as the pseudo-secular approximation, leaving, in addition to the secular terms, also the operators $\hat{I}_+\hat{S}_z$ and $\hat{I}_-\hat{S}_z$ to be considered. In fact, these terms are of key importance as they are actually responsible for the DNP effect. Furthermore, it is safe to disregard the flip-flop terms $\hat{I}_+\hat{S}_-$ and $\hat{I}_-\hat{S}_+$ due to the large energy difference between nuclear and electronic Zeeman interactions.

In the rotating frame of the microwave irradiation only (instead of the doubly rotating frame, in order to be consistent with the pseudo-secular instead of the secular approximation) we can rewrite the spin Hamiltonian according to:
\begin{equation}
\tilde{\hat{H}}_{0}=\frac{1}{2}
\begin{pNiceMatrix}[first-row,last-col=5]
 \bra{}{\alpha\alpha}   & \bra{}{\alpha\beta}   &\bra{}{\beta\alpha}    &\bra{}{\beta\beta}& \\
\Delta\omega_e+\omega_n +\frac{1}{2}A\omega_{d}^{en} & C\omega_{d}^{en} &  0 &0 &\ket{}{\alpha\alpha}\\
 D\omega_{d}^{en}  &\Delta\omega_e-\omega_n-\frac{1}{2}A\omega_{d}^{en}  &0 & 0&\ket{}{\alpha\beta}\\
  0  &0 &-\Delta\omega_e+\omega_n-\frac{1}{2}A\omega_{d}^{en} &- C\omega_{d}^{en} &\ket{}{\beta\alpha}\\
 0  & 0 &-D\omega_{d}^{en}& -\Delta\omega_e-\omega_n+\frac{1}{2}A\omega_{d}^{en} &\ket{}{\beta\beta}
\end{pNiceMatrix}
.
\label{eq:pseudo}
\end{equation}
Here we introduced $\Delta\omega_e=\omega_{e}-\omega_{\mu w}$, the off-resonance frequency of the electron spins with respect to the microwave frequency. We shall make an additional approximation, for aesthetic purposes only, and neglect the secular part of the dipolar coupling (the through space dipolar coupling of a $^{17}$O nucleus to a spin 1/2 electron at a distance of 2\AA~is 1~MHz, while its Larmor frequency at 9.4~T is 54~MHz):
\begin{equation}
\tilde{\hat{H}}_{0}=\frac{1}{2}
\begin{pNiceMatrix}[first-row,last-col=5]
 \bra{}{\alpha\alpha}   & \bra{}{\alpha\beta}   &\bra{}{\beta\alpha}    &\bra{}{\beta\beta}& \\
\Delta\omega_e+\omega_n  & C\omega_{d}^{en} &  0 &0 &\ket{}{\alpha\alpha}\\
 D\omega_{d}^{en}  &\Delta\omega_e-\omega_n  &0 & 0&\ket{}{\alpha\beta}\\
  0  &0 &-\Delta\omega_e+\omega_n &- C\omega_{d}^{en} &\ket{}{\beta\alpha}\\
 0  & 0 &-D\omega_{d}^{en}& -\Delta\omega_e-\omega_n &\ket{}{\beta\beta}
\end{pNiceMatrix}
.
\end{equation}

Finally, we can write now the total Hamiltonian as:
\begin{equation}
\tilde{\hat{H}}_{tot}=\tilde{\hat{H}}_{0}+\tilde{\hat{H}}_{\mu w}=\frac{1}{2}
\begin{pNiceMatrix}[first-row,last-col=5]
 \bra{}{\alpha\alpha}   & \bra{}{\alpha\beta}   &\bra{}{\beta\alpha}    &\bra{}{\beta\beta} &\\
\Delta\omega_e+\omega_n & C\omega_{d}^{en} &  \omega_1 &0 &\ket{}{\alpha\alpha}\\
 D\omega_{d}^{en}  &\Delta\omega_e-\omega_n  &0 & \omega_1&\ket{}{\alpha\beta}\\
 \omega_1  &0 &-\Delta\omega_e+\omega_n &- C\omega_{d}^{en} &\ket{}{\beta\alpha}\\
 0  & \omega_1 &-D\omega_{d}^{en}& -\Delta\omega_e-\omega_n &\ket{}{\beta\beta}
 \end{pNiceMatrix}
.
\label{eq:Htot}
\end{equation}
        
\subsection{The Solid Effect}

\subsubsection{The eigenstate representation of the Hamiltonian}

The SE DNP mechanism requires irradiation of the DQ or ZQ transitions, as shown in Fig.~(\ref{fg:enhpop}~a). However, from the Hamiltonian in equation~(\ref{eq:Htot}) one can see that the microwave terms do not directly connect either DQ or ZQ transitions. Instead, it is the combined effect of the microwave irradiation and the pseudosecular term of the dipolar interaction that will make the DQ or ZQ irradiation effective. To demonstrate this, it is useful to write the Hamiltonian $\tilde{\hat{H}}_{0}$ in its eigenstate representation, diagonalizing it:
\begin{equation}
\begin{split}
\tilde{\hat{\Lambda}}_{0}&=\hat{D}^{-1}\tilde{\hat{H}}_{0}\hat{D}\\
&\approx \frac{1}{2}
\begin{pNiceMatrix}[first-row,last-col=5]
 \bra{}{\alpha\alpha}'   & \bra{}{\alpha\beta}'   &\bra{}{\beta\alpha}'   &\bra{}{\beta\beta}' &\\
\Delta\omega_e+\omega_n-\frac{\left(|C|\omega_{d}^{en}\right)^2}{2\omega_n} &   0 & 0 &0 &\ket{}{\alpha\alpha}'\\
0 &\Delta\omega_e-\omega_n +\frac{\left(|C|\omega_{d}^{en}\right)^2}{2\omega_n}  &0 & 0&\ket{}{\alpha\beta}'\\
 0  &0 &-\Delta\omega_e+\omega_n-\frac{\left(|C|\omega_{d}^{en}\right)^2}{2\omega_n}  &0 &\ket{}{\beta\alpha}'\\
 0  & 0 &0& -\Delta\omega_e-\omega_n+\frac{\left(|C|\omega_{d}^{en}\right)^2}{2\omega_n}  &\ket{}{\beta\beta}'
 \end{pNiceMatrix}
\end{split}
,
\label{eq:Hdiag}
\end{equation}
with the eignestates related to the Zeeman basis according to:
\begin{equation}
\begin{split}
    \bra{}{\alpha\alpha}'&=c\bra{}{\alpha\alpha}+s\bra{}{\alpha\beta},\\
    \bra{}{\alpha\beta}'&=c\bra{}{\alpha\beta}-s\bra{}{\alpha\alpha},\\
    \bra{}{\beta\alpha}'&=c\bra{}{\beta\alpha}+s\bra{}{\beta\beta},\\
    \bra{}{\beta\beta}'&=c\bra{}{\beta\beta}-s\bra{}{\beta\alpha},
\end{split}
\end{equation}
with
\begin{equation}
c\approx1\quad \text{and} \quad
s\approx\frac{|C|\omega_{d}^{en}}{2\omega_n}.
\end{equation}

\begin{figure*}
\begin{center}
\includegraphics[scale=0.75]{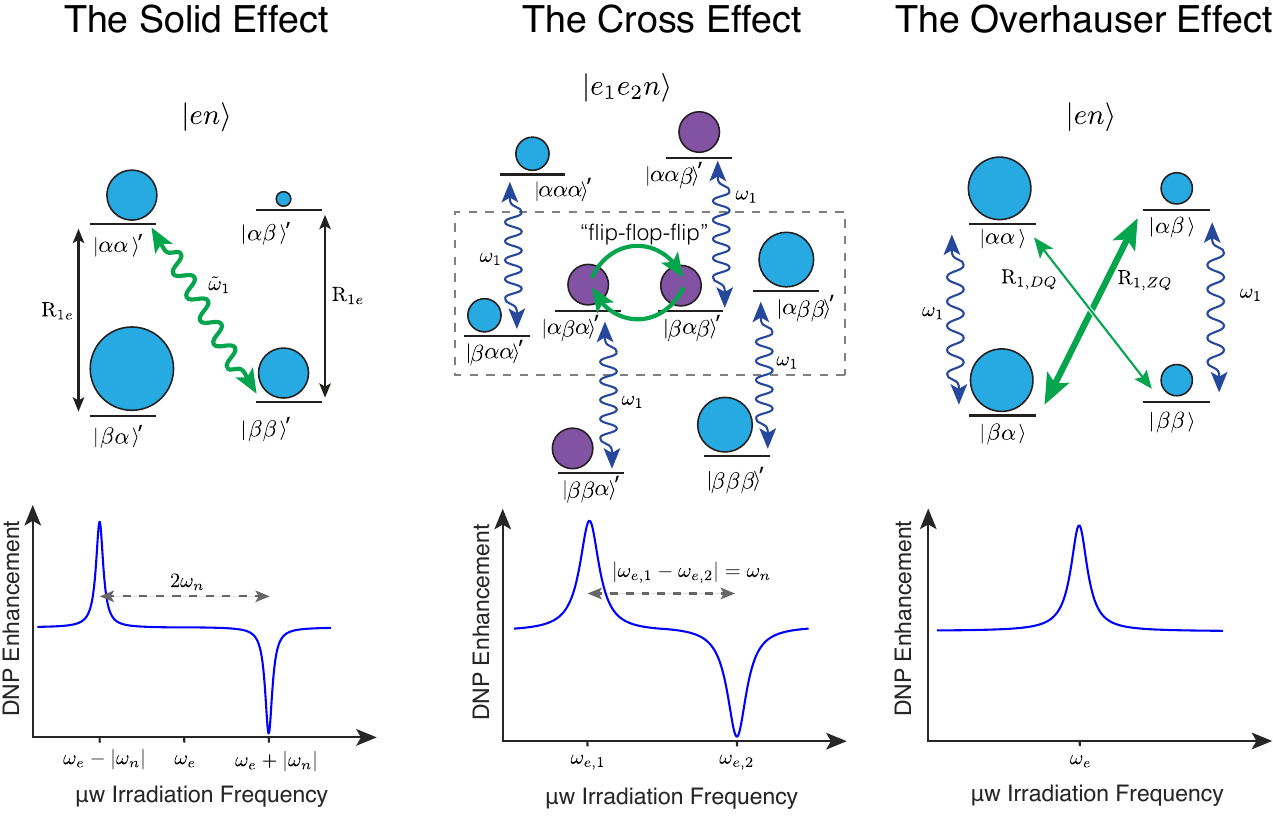}
\end{center}	
\caption{{\bf Summary of the three main DNP mechanisms.} 
$Left$: The solid effect requires one electron coupled to one nucleus. Saturation of either the zero- or double quantum transition leads to positive and negative DNP enhancement, respectively. Maximum of positive and negative enhancements are therefore separated by twice the nuclear Larmor frequency and centred at the electron Larmor frequency.
$centre$: Cross effect can occur in a system of two coupled electron spins where at least one of them is further coupled to a nuclear spin and there is a degeneracy between two energy levels according to: $|\omega_{e,1}-\omega_{e,2}|=|\omega_n|$. Saturation of either of the electron single quantum transitions will lead to a difference in electron polarization, this polarization difference can give raise to nuclear hyperpolarization via the  "flip-flop-flip" transition of the degenerate states.
$Right$: The Overhauser effect requires one electron coupled to one nucleus, saturation of the electron single quantum transition and an imbalance in the $R_{1,DQ}$ and $R_{1,ZQ}$ relaxation rates. For $R_{1,ZQ}>\text{R}_{1,DQ}$, and a positive nuclear gyromagnetic ratio, the enhancement is positive. Maximum enhancement is obtained when irradiating on resonance with the electron single quantum transition.
Note that the DNP profiles are shown as a function of the microwave irradiation frequency, and at a constant magnetic field, whereas in most experimental setups actually the microwave irradiation frequency is constant and the magnetic field is swept.}
\label{fg:DNPmechanisms}
\end{figure*}

In order to express the microwave Hamiltonian in the same basis, we have to transform it with the same diagonalization matrix $\hat{D}$, leading to:
\begin{equation}
\tilde{\hat{H}}_{\mu w}^{\Lambda}\approx\frac{1}{2}
\begin{pNiceMatrix}[first-row,last-col=5]
 \bra{}{\alpha\alpha}'   & \bra{}{\alpha\beta}'   &\bra{}{\beta\alpha}'    &\bra{}{\beta\beta}' &\\
0& 0 &  \omega_1 &\omega_1\frac{|C|\omega_{d}^{en}}{\omega_n} &\ket{}{\alpha\alpha}'\\
0 &0 &-\omega_1\frac{|C|\omega_{d}^{en}}{\omega_n} & \omega_1&\ket{}{\alpha\beta}'\\
 \omega_1  &-\omega_1\frac{|C|\omega_{d}^{en}}{\omega_n} &0 &0 &\ket{}{\beta\alpha}'\\
 \omega_1\frac{|C|\omega_{d}^{en}}{\omega_n}  & \omega_1 &0& 0 &\ket{}{\beta\beta}'
 \end{pNiceMatrix}
.
\label{eq:HMWnew}
\end{equation}
This matrix shows the presence of elements connecting the formally forbidden DQ and ZQ transitions. The off-diagonal elements of the Hamiltonian become effective when differences between the diagonal elements of the corresponding subspace are small. This occurs at the SE conditions: $\Delta\omega_e\approx-\omega_n$ for the DQ transition and at  $\Delta\omega_e\approx\omega_n$ for the ZQ transition (see equation~(\ref{eq:Hdiag})), leading to positive and negative enhancements, respectively (Fig.~\ref{fg:DNPmechanisms}). When the microwave frequency satisfies $\Delta\omega_e\approx-\omega_n$ this means that the irradiation frequency is at $\omega_e+\omega_n$ (remembering that for a positive $\gamma_n$, the nuclear Larmor frequency $\omega_n$ is negative). Finally, at the SE condition, the effective nutation frequency of the DQ and ZQ transitions is:
        \begin{equation}
\tilde{\omega}_1\approx\omega_1\frac{|C|\omega_{d}^{en}}{\omega_n}.
\label{eq:effMW}
\end{equation}
The effective irradiation amplitude is proportional to the strength of the dipolar coupling scaled by the nuclear Larmor frequency. Therefore, the effective nutation frequency becomes weaker at higher magnetic fields. Note that throughout these equations we have replaced the pseudosecular coefficients $C=D^*$ by their modulus $|C|$, which is the relevant quantity.

        \subsubsection{The steady-state polarization}
                    \label{subsec:SSpolarization}
The saturation efficiency of the DQ and ZQ transitions upon microwave irradiation (Fig.~(\ref{fg:enhpop})) is given by:\cite{jpcl_11_5439_2020}
\begin{equation}
\Delta p_{DQ/ZQ}=\frac{\Delta p_{DQ/ZQ,eq}}{1+\tilde{\omega}_1^2\left[R_{2e}(2R_{1,DQ}+2R_{1})\right]^{-1}}.
\label{eq:DNPsat}
\end{equation}
In the Appendix we show how to derive this equation from the Bloch equations that describe saturation. In the derivation we assume that the population difference between electronic spin states is small (high-temperature approximation), but much larger than that between nuclear spin states. Looking at the equation we see that the saturation efficiency depends on the effective nutation frequency $\tilde{\omega}_1$ as well as on the transverse relaxation rate of the electron $R_{2e}$ and the longitudinal relaxation rates of the nucleus $R_{1}$ and of the DQ/ZQ transition $R_{1,DQ}\approx R_{1,ZQ}$. In principle, nuclear and electron relaxation times can be estimated experimentally, and in the preceding subsection we have derived an expression for the effective nutation frequency of those transitions. Thus, the parameter in equation~(\ref{eq:DNPsat}) that has yet to be estimated is $R_{1,DQ}$. In the following, an approach, proposed by Hovav et al.\cite{jmr_207_176_2010} is presented that estimates $R_{1,DQ}$ by applying the same diagonalization procedure to the relaxation operators.

As previously mentioned, longitudinal relaxation is caused by fluctuations of local magnetic fields, more specifically, the components of local fields perpendicular to the quantization axis. For simplicity, assuming that the fluctuating component is pointing along the $x$-axis, one could write:
        \begin{equation}
R_1\propto\langle B_x^2 \rangle J(\omega),
\end{equation}
where B$_x$ is the fluctuating local field and $J(\omega)$ is the spectral density function at the transition frequency. We have  shown how an operator along $x$ acting on the electron spin will also affect the DQ and ZQ transitions of the coupled spin system, by rotating $\hat{S}_x$ into the eigenstate basis. In the same way, we can estimate the effect of the fluctuating B$_x$ field on the longitudinal relaxation of the DQ and ZQ transitions leading to:
        \begin{equation}
R_{1,DQ}=R_{1e}\left(\frac{|C|\omega_d^{en}}{\omega_n}\right)^2.
\label{eq:R1dq}
\end{equation}
The same rationale could be applied to the nuclear relaxation rate, but since it is much slower than the electron relaxation rate, we can neglect its contribution to $R_{1,DQ}$.

Finally, from simple considerations as shown in Fig.~(\ref{fg:enhpop} b) one can show that the nuclear polarization enhancement is related to the saturation efficiency according to:
        \begin{equation}
P_{n}/P_{e}\approx\frac{P_{e}-\Delta p_{DQ}}{P_{e}},
\label{eq:DNPenhancement}
\end{equation}
where $P_n$ and $P_e$ are the nuclear and electron polarization values at steady state. The only assumptions here are that the electron relaxation rate is much faster than any other rate affecting the system, such that:
        \begin{equation}
\frac{p_{4}^{eq}}{p_{2}^{eq}}=\frac{p_{4}^{ss}}{p_{2}^{ss}}=\frac{p_{3}^{eq}}{p_{1}^{eq}}=\frac{p_{3}^{ss}}{p_{1}^{ss}}=\text{const},
\label{eq:T1epopratio}
\end{equation}
and that the total population at thermal equilibrium and the steady-state is conserved:
        \begin{equation}
p_{1}^{eq}+p_{2}^{eq}+p_{3}^{eq}+p_{4}^{eq}=p_{1}^{ss}+p_{2}^{ss}+p_{3}^{ss}+p_{4}^{ss}.
\end{equation}

Equation~(\ref{eq:DNPenhancement}) shows that for full saturation ($\Delta p_{DQ}=0$) the nuclear steady-state polarization equals the electron polarization $P_n=P_e$, the theoretical limit of the DNP enhancement, while in the absence of microwave irradiation $\Delta p_{DQ}=P_{e,eq}-P_{n,eq}$ (again, see Fig.~(\ref{fg:enhpop})) and the nuclear steady-state polarization $P_n$ simply remains $P_{n,eq}$.

   \subsection{The cross effect}
    
The CE DNP mechanism requires three spins, two coupled electrons and one nucleus coupled to at least one of the electrons.\cite{jmr_214_29_2012} DNP can occur when the difference in the resonance frequency of the two electrons equals the nuclear Larmor frequency, the CE condition: 
        \begin{equation}
|\omega_{e,1}-\omega_{e,2}|=|\Delta \omega_{e12}|=|\omega_n|.
\end{equation}
When this condition is fulfilled there will be a degeneracy of two energy levels $\bra{}{e_1e_2n}$, either $\bra{}{\alpha\beta\alpha}\leftrightarrow\bra{}{\beta\alpha\beta}$ or $\bra{}{\beta\alpha\alpha}\leftrightarrow\bra{}{\alpha\beta\beta}$. This degeneracy results in a very efficient population transfer and can lead to nuclear hyperpolarization when a population difference between both degenerate levels is created, for instance through selective saturation of the single quantum transition of one of the electrons, as illustrated in Fig.~(\ref{fg:DNPmechanisms}).

To describe the spin physics of the CE mechanism it is useful to follow a treatment in analogy to the one presented for the SE mechanism.\cite{jcp_137_084508_2012} For that purpose we will need to include in the Hamiltonian the Zeeman interaction of the second electron as well as the electron-electron dipolar coupling, while for simplicity we will neglect the dipolar coupling between the nucleus and the second electron. Further, we will neglect the diagonal terms of the dipolar coupling (the first term in the dipolar alphabet, equation~(\ref{eq:dipalph})). Thus, the relevant terms of the Hamiltonian for the  energy levels highlighted within the dashed box in  Fig.~(\ref{fg:DNPmechanisms}) are given by:

\begin{equation}
\tilde{\hat{H}}_{0}=\frac{1}{2}
\begin{pNiceMatrix}[first-row,last-col=5]
 \bra{}{\alpha\beta\alpha}   & \bra{}{\alpha\beta\beta}   &\bra{}{\beta\alpha\alpha}    &\bra{}{\beta\alpha\beta}& \\
\Delta\omega_e+\omega_n  & C\omega_{d}^{en} &   B_{ee}\omega_d^{ee}  &0 &\ket{}{\alpha\beta\alpha}\\
 D\omega_{d}^{en}  &\Delta\omega_e-\omega_n &0 &  B_{ee}\omega_d^{ee} &\ket{}{\alpha\beta\beta}\\
  B_{ee}\omega_d^{ee}  &0 &-\Delta\omega_e+\omega_n &- C\omega_{d}^{en} &\ket{}{\beta\alpha\alpha}\\
 0  &  B_{ee}\omega_d^{ee}  &-D\omega_{d}^{en}& -\Delta\omega_e-\omega_n &\ket{}{\beta\alpha\beta}
\end{pNiceMatrix}
.
\end{equation}

In a first step, we can diagonalize the Hamiltonian containing the Zeeman interactions and the electron-nuclear dipolar interaction only, obtaining a new basis composed of mixed Zeeman states (in full analogy to the SE derivations, thus not shown here). And in a second step we rewrite the flip-flop term of the electron-electron coupling in this new basis:

\begin{equation}
\tilde{\hat{H}}_{B}^\Lambda=\frac{1}{2}
\begin{pNiceMatrix}[first-row,last-col=5]
 \bra{}{\alpha\beta\alpha}'   & \bra{}{\alpha\beta\beta}'   &\bra{}{\beta\alpha\alpha}'    &\bra{}{\beta\alpha\beta}'& \\
0  & 0 &   B_{ee}\omega_d^{ee}  &B_{ee}\omega_d^{ee}\frac{|C|\omega_d^{en}}{\omega_n} &\ket{}{\alpha\beta\alpha}'\\
 0  &0 &B_{ee}\omega_d^{ee}\frac{|C|\omega_d^{en}}{\omega_n} &  B_{ee}\omega_d^{ee} &\ket{}{\alpha\beta\beta}'\\
  B_{ee}\omega_d^{ee}  &B_{ee}\omega_d^{ee}\frac{|C|\omega_d^{en}}{\omega_n} &0 &0 &\ket{}{\beta\alpha\alpha}'\\
B_{ee}\omega_d^{ee}\frac{|C|\omega_d^{en}}{\omega_n}  &  B_{ee}\omega_d^{ee}  &0& 0 &\ket{}{\beta\alpha\beta}'
\end{pNiceMatrix}
.
\end{equation}

Comparing this Hamiltonian with equation~\ref{eq:HMWnew} shows the equivalence of the roles of the dipolar coupling in the CE and the microwave irradiation in the SE DNP mechanisms. However, it is important to highlight that the CE DNP still requires irradiation with microwave to create a population difference. But, unlike in the SE, it suffices to irradiate an allowed single-quantum transition of one of the electrons. Full saturation of this transition will lead to the maximum theoretical enhancement, $\gamma_e/\gamma_n$. Since the nutation frequency does not get scaled by $\omega^{en}_d/\omega_n$, the microwave power requirement for the cross effect is significantly lower.

Interestingly, saturation of the electron transition and matching of the CE condition do not have to occur at the same time. This is exploited in CE MAS DNP, where due to their anisotropy, the electron transition frequencies are modulated by the sample spinning, eventually passing through orientations that match either the microwave irradiation frequency, the cross effect condition or a dipolar condition, where the frequency of both electrons is the same.\cite{jcp_137_084508_2012,jmr_224_13_2012}
During these so-called `rotor events', or energy level anti-crossings, changes in the populations of the different energy levels can occur.
The efficiency of these changes can be analyzed in terms of the Landau-Zener formalism. The relevant parameters are the size of the involved dipolar or microwave interactions and the rate of change of the anisotropic interactions (which depends on the size of the anisotropy and on the spinning speed). 
Since the contributions are additive, they can lead to large nuclear polarization increment over the course of many rotor periods, even for a very low efficiency of individual events, .\cite{jmr_258_102_2015}  Because of that, and thanks to the development of nitroxide biradicals tailored for exactly this purpose, the CE mechanism leads to the largest signal enhancements in recent high field MAS DNP applications.\cite{jacs_126_10844_2004,jacs_128_11385_2006,ac_54_11770_2015,pccp_22_3643_2020,jpcl_11_8386_2020,cs_11_2810_2020}

Until now the CE has not been very efficient in MI-DNP. The main two reasons for this are:\cite{jpcc_127_4759_2023} ($1$) MI-DNP has been most successful in crystalline materials, when paramagnetic dopants are doped into the sample of interest, it is most likely that they will occupy a given preferred crystallographic site. In a powder containing micron-sized particles, pairs of electron spins are most likely to be in the same crystallite, therefore, their interaction tensors will have the same orientation at any time during spinning, so that the anisotropy of the g- and ZFS-interactions cannot be a source of an energy difference. And ($2$) $T_{1e}$ relaxation times in paramagnetic metal ions reported so far are considerably shorter compared to state-of-the-art nitroxide radicals. When the relaxation time is short compared to the timescale of a rotor period, the changes in populations caused by the rotor events will not be retained until the next event to form a constructive cascade, instead the system will relax back to the Boltzmann equilibrium. The most common commercial DNP instruments can reach MAS speeds of about 10~kHz at 100~K, which corresponds to a rotor period of 100~$\mu$s, while on the other hand, reported $T_{1e}$ of metal ions used for MI-DNP are in the order of few $\mu$s.\cite{jpcc_124_7082_2020,jmr_336_107143_2022}

   \subsection{The Overhauser effect}

The Overhauser effect is fundamentally different from the solid and cross effect as it is not a coherent effect. Instead, this DNP mechanism is driven by relaxation, specifically, by differences in the ZQ and DQ cross-relaxation rates that can lead to nuclear hyperpolarization if the electron SQ transition is saturated (Fig.~\ref{fg:DNPmechanisms}).  The relaxation behaviour of a pair of coupled spins $I$ and $S$ due to the modulation of their coupling is described by the Solomon equations.\cite{pr_99_559_1955} According to these equations the evolution of the longitudinal magnetization of spin $I$ is given by:
 \begin{equation}
\frac{d\langle\hat{I}_z\rangle}{dt}=-R_{auto}\left(\langle\hat{I}_z\rangle-\langle\hat{I}_z\rangle_0\right)-R_{cross}\left(\langle\hat{S}_z\rangle-\langle\hat{S}_z\rangle_0\right),
\end{equation}
where $R_{auto}$ and $R_{cross}$ are the auto- and cross-relaxation rates. In terms of spectral density functions, the rates for the case of relaxation mediated by through-space dipolar couplings is given by  (after averaging over all angles in equation~\ref{eq:T1PRE_c}):
 \begin{equation}
\begin{split}
R_{auto}^{d}=&\frac{2}{15}(S(S+1)) \left(\omega_d^{en}\right)^2\left(J(\omega_n-\omega_e)+3J(\omega_n)+6J(\omega_n+\omega_e)\right),\\ 
R_{cross}^{d}=& -\frac{2}{15}(I(I+1))
\left(\omega_d^{en}\right)^2\left(-J(\omega_n-\omega_e)+6J(\omega_n+\omega_e)\right).
\end{split}
\label{eq:autocross1}
\end{equation}
For relaxation mediated by the Fermi contact couplings, the rates are: 
 \begin{equation}
\begin{split}
R_{auto}^{FC}=&\frac{2}{3} \left(A_{FC}\right)^2\left(J(\omega_n-\omega_e)\right),\\ 
R_{cross}^{FC}=& -\frac{2}{3}\left(A_{FC}\right)^2\left(J(\omega_n-\omega_e)\right),
\end{split}
\label{eq:autocross2}
\end{equation}
where $J(\omega)=\tau_c/(1+\omega^2\tau_c^2)$ are the spectral densities introduced in previous sections. The implication of the cross-relaxation terms is that magnetization will be transferred from one spin to the other. This transfer can lead to a polarization enhancement of spin $I$ by up to the ratio of gyromagnetic ratios $\gamma_S/\gamma_I$, when the single quantum transition of spin $S$ is fully saturated. Overall, Overhauser enhancement requires that three conditions be fulfilled:\cite{jmr_264_78_2016}
\begin{itemize}
     \item A significant saturation of the single-quantum transitions of the electron S is achieved. This is quantified by the saturation factor $s$ which has a range $0<s<1$.    
     \item The relaxation rates given in the previous equations are the main source of relaxation. These rates ares largest when the correlation time is close to the inverse of the electron Larmor frequency, which is in the picosecond time scale. If other relaxation mechanisms are more efficient, they will counteract the build-up of hyperpolarization. The leakage factor $f$ quantifies this relation and can also take values between $0<f<1$. 
    \item And finally, the ratio $R_{cross}/R_{auto}=\zeta$ has to be large. Analysis of equations~(\ref{eq:autocross1}) and (\ref{eq:autocross2}) reveals that $\zeta$ has a constant value of $\zeta=-1$ for a hypothetical case of purely contact driven relaxation, while it has a range $0<\zeta<0.5$ when relaxation is only due to the through space-dipolar interaction (the signs assume a positive nuclear gyromagnetic ratio). Note that for high magnetic fields, as $\omega_e\tau_c$ becomes larger than unity, the single quantum nuclear spectral density $J(\omega_n)$, present only in the dipolar case, will eventually become the dominant contribution, driving $\zeta$ to 0. Due to the opposite signs of $\zeta$ in through-space dipolar and FC driven cross relaxation, the enhancement from a combination of both mechanisms will always lead to an attenuation of the polarization.
\end{itemize}
The enhancement factor will depend on the efficiency of the three conditions according to: 
 \begin{equation}
\epsilon=\zeta f s \frac{\gamma_e}{\gamma_n}.
\end{equation}

In recent years several cases of efficient Overhauser DNP at high magnetic fields have been reported, including experiments in the liquid,\cite{jacs_131_6090_2009,nc_9_676_2017,ac_58_1402_2019,nc_12_6880_2021} and in solid state, not only in metals\cite{nc_11_2224_2020} but also in insulating frozen matrices containing organic radicals.\cite{jcp_141_064202_2014,jpcl_8_2137_2017,jpcs_125_867_2021,jpcl_13_4000_2022}
However, no reports of Overhauser enhancements at high fields using paramagnetic metal ions have been reported so far, to our knowledge. The reasons for this difficulty have been discussed based on theoretical considerations for both solids\cite{pr_98_1729_1955} and liquids\cite{jbnmr_58_239_2014} and are mostly related to the absence of motions on an adequate time scale, thus a low leakage factor.

\section{Metal Ions DNP}
    \label{sec:MIDNP}

In the preceding sections we have seen how EPR spectroscopy can assist in characterizing paramagnetic metal ions introduced as dopants, we discussed how their presence will affect the NMR properties of the sample, and lastly we introduced the most common DNP mechanisms for high field MAS DNP applications. Throughout those sections we emphasized the conditions under which MI-DNP can be expected to be viable. In this section we will focus on the peculiarities of MI-DNP mainly compared to the more common MAS DNP approach using nitroxide radicals. Most importantly, as the lattice affects the properties of the dopant, the polarizing agents will behave differently in every sample and require  characterization. In this regard, DNP field sweep profiles play a fundamental role (see subsection~\ref{subsec:DNPsweep}). The fact that the polarizing agent is located within the sample opens some exciting possibilities, as will be treated in subsection~\ref{subsec:endogenous}. Finally, some more general aspects will cover the diffusion of the polarization throughout the sample  and the role of paramagnetic relaxation enhancement, focusing on low sensitivity nuclei in unprotonated solids, where spin diffusion is inefficient (subsection~\ref{subsec:polvsRelax}).

    \subsection{DNP field sweep profiles}
        \label{subsec:DNPsweep}
A necessary step for DNP is identifying the best conditions of microwave frequency and magnetic field that will give rise to optimum polarization transfer. This is achieved by acquiring a DNP field sweep profile, where NMR spectra are either recorded at various magnetic fields under microwave irradiation at constant frequency and power, or, depending on the instrument, at constant field and varying microwave frequency. In our discussions we will focus on the sweeps of the magnetic field, as this is the method used in many commercially available instruments. The integrated intensity of the acquired NMR spectra is plotted against the magnetic field. Field sweep profiles are highly informative, as they not only depend on the dominant DNP mechanism, but primarily reflect the EPR properties of the paramagnetic centre. In many cases the features of the DNP sweep profile simply $mirror$ the EPR spectrum, and it is possible to deconvolute the contributions from different DNP mechanisms that act simultaneously.\cite{pccp_14_5729_2012} However, this is an approximation and additional information on the electron-electron coupling network might be required for a full understanding of the observed DNP sweep.\cite{pccp_17_6053_2015,jpcl_11_3718_2020}

            \begin{figure*}
\begin{center}
\includegraphics[scale=1]{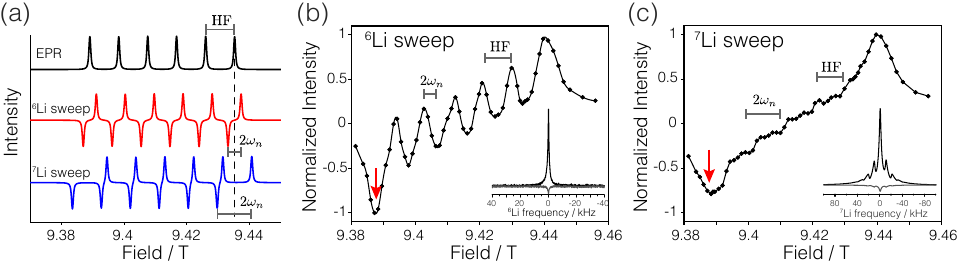}
\end{center}	
\caption{{\bf MI-DNP field sweep profiles.} (a) Simulated Mn(II) EPR spectrum and corresponding $^6$Li and $^7$Li field sweep profiles, using a hyperfine coupling $A_{FC}=259$~MHz and a microwave irradiation frequency of 263.601~GHz. At this field twice the nuclear Larmor frequency is $2\omega_n(^6\text{Li})=118$~MHz and $2\omega_n(^7\text{Li})=311$~MHz, for $^6$Li and $^7$Li, respectively. In (b) and (c)  the experimental $^6$Li and $^7$Li sweep profiles in LiMg$_{1-x}$Mn$_x$PO$_4$  with $x=0.005$ obtained at 100 K, under MAS, are shown. The insets show the spectra obtained with and without microwave irradiation, at the field position indicated with red arrows in the sweeps. (b) and (c) are reproduced with permission from Ref.~\cite{jmr_336_107143_2022}.}
\label{fg:sweeps}
\end{figure*}

In MI-DNP, the presence of strong interactions such as ZFS and hyperfine couplings to the metal's nucleus will impact the appearance of the EPR spectrum, and consequently of the DNP field sweep profile. Fig.~(\ref{fg:sweeps} a) shows a simulated EPR spectrum of a Mn(II) ion at a microwave frequency of 263.601~GHz,  assuming only isotropic hyperfine couplings to $^{55}$Mn. The EPR spectrum shows a characteristic sextet due to the coupling with the nuclear spin $I=5/2$ of $^{55}$Mn. Solid effect DNP will result in a positive and a negative enhancement for the DQ and ZQ transitions respectively, separated by $\pm\omega_n$ from each electron transition. Consequently, the DNP field sweep profile in this case results in 6 regions with positive and 6 regions with negative signal enhancements, as shown in the figures below for $^6$Li and $^7$Li. Note that the sweep profile will have a different appearance according to the relative sizes of $2\omega_n$ and $A_{FC}$.

Fig.~(\ref{fg:sweeps} b and c) shows experimental field sweep profiles for $^6$Li and $^7$Li in LiMg$_{1-x}$Mn$_x$PO$_4$. While in the $^6$Li case all expected features are nicely resolved, in the $^7$Li sweep, due to the larger Larmor frequency, most of the positive and negative enhancement lobes partially cancel each other, so that only the outer most lobes are clearly resolved. Note that for better visualization in Fig.~(\ref{fg:sweeps} a),  the simulated EPR lines are significantly narrower than they are in the real experiments. 

Clearly, the optimum field position for maximal DNP enhancement will strongly depend on the EPR properties of the metal ion in each individual sample. Changes in the g-value, $A_{FC}$, and ZFS will affect the outcome. From a practical point of view, this means that unlike in nitroxide formulations, where the sample of interest has little effect on the EPR properties of the polarizing agent, in MI-DNP each sample requires optimization of the static magnetic field to obtain maximum signal enhancement. Of course, a thorough characterization of the sample with EPR prior to the DNP measurements will facilitate this procedure. However, small uncertainties in field and microwave frequency of the EPR at low magnetic fields can have large impact at the higher fields at which DNP is performed. Furthermore, it is not always straightforward to disentangle the role of all interactions on line shape and position, which might impede accurate projections to higher magnetic fields.

A further important aspect that becomes apparent from the Mn(II) field sweep profile shown is that not all electron spins will contribute to the DNP effect simultaneously. The splitting of the EPR line into a sextet reduces the available polarization to 1/6. In addition, since Mn(II) has an electronic spin 5/2, only the CT is likely to contribute significantly to the signal enhancement, so that another factor of 3 in sensitivity reduction has to be considered.\cite{jacs_133_5648_2011}

    \subsection{Exogenous and endogenous DNP}
            \label{subsec:endogenous}
\begin{figure*}
\begin{center}
\includegraphics[scale=1]{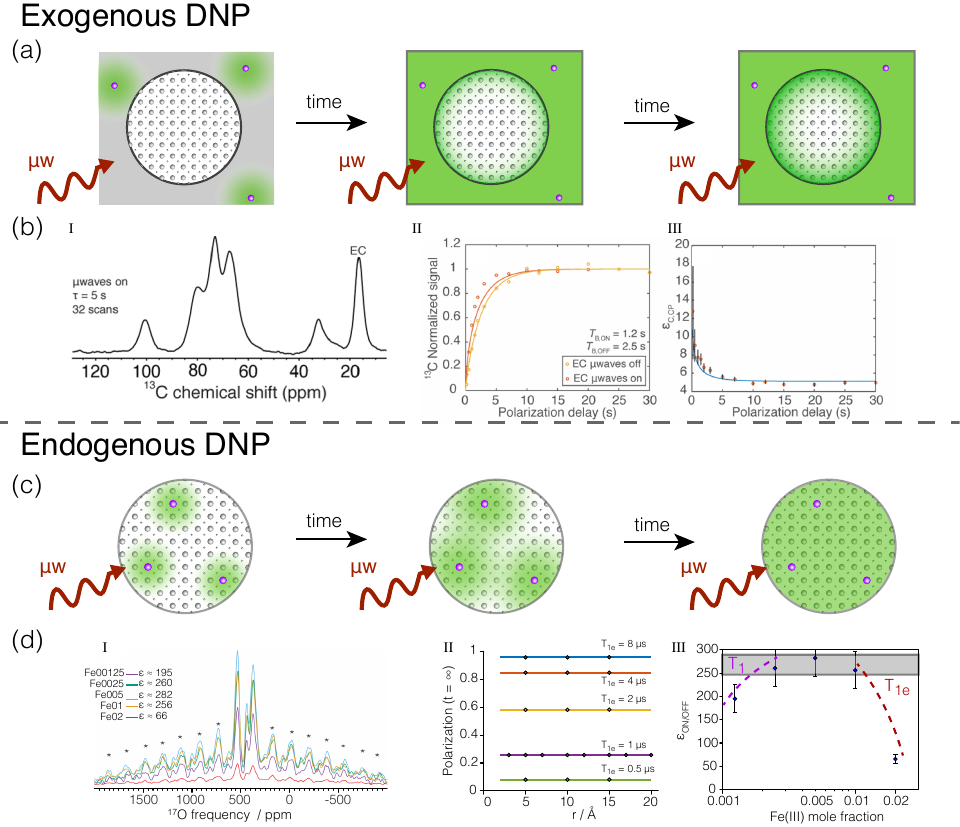}
\end{center}	
\caption{{\bf Main differences between the exogenous and endogenous MAS DNP approaches.} (a) Schematic representation of an exogenous DNP formulation. The sample is soaked in a solvent containing the polarizing agent. Upon cooling, the solvent forms a glassy matrix. Hyperpolarization reaches the sample either through direct polarization or, more commonly, via protons from the solvent. For the latter, proton nuclei in the vicinity of the polarizing agent are hyperpolarized and subsequently the polarization spreads through the solvent via efficient spin diffusion, eventually reaching the sample surface. This approach is therefore highly surface selective. (b) $I$ to $III$ show experimental results of $^{13}$C CPMAS DNP in ethyl cellulose nanoparticles impregnated with TCE/MeOH containing the bCTbK nitroxide biradical. Since the enhancement is most prominent for fast relaxing surface sites, the overall enhancement factor decreases with increasing polarization delay.  Figure reproduced with permission from Ref.~\cite{jpcc_121_15993_2017}; (c) In endogenous DNP the polarizing agent is introduced inside the sample of interest as a dopant. When PRE from the polarizing agent is the main source of relaxation the steady state enhancements are distance independent and homogeneous enhancements throughout the sample are obtained. (d) $^{17}$O direct MAS DNP in Li$_4$Ti$_5$O$_{12}$ doped with Fe(III) shows a constant enhancement over a large dopant concentration range. Figure adapted with permission from Ref.~\cite{jpcl_11_5439_2020}.  }
\label{fg:endoexo}
\end{figure*}

The strongest distinction between the MI-DNP approach and the use of organic radicals is that the source of polarization can be introduced into the sample of interest itself through doping. Let us first review the workflow in an exogenous DNP experiment, so as to later highlight the differences. Fig.~(\ref{fg:endoexo}~a) shows schematically a DNP experiment using exogenous radicals on a non-soluble sample. Most current applications of MAS DNP on materials follow the exogenous approach.\cite{ssnmr_101_116_2019}  This consists in surrounding the sample with a frozen solution containing organic radicals which serve as polarizing agents. Significant efforts are devoted to optimize the formulation, including the nature of the radicals, mostly bi-radicals, their concentration, as well as the solvent itself. Upon microwave irradiation the polarization is generally transferred from the electron spins to the protons on the radical molecule and/or to solvent protons in the immediate proximity. Subsequently, spin diffusion through the proton spin bath spreads the polarization throughout the frozen solvent reaching the surface of the sample and, if the sample contains protons, eventually penetrating the sample. In order to enhance heteronuclei the polarization can be either transferred from the radicals directly to the nuclei of interest, or more commonly,  via the protons: after hyperpolarizing the proton spin bath, a cross-polarization step can be used to transfer the polarization from the protons to the nuclei of interest.

In endogenous DNP on the other hand, the polarizing agents are part of the structure of the material of interest. While some materials might intrinsically have some source of unpaired electrons with favourable DNP properties,\cite{pnmrs_17_33_1985,f_66_876_1987,jpcc_119_19272_2015,jpcc_2022,prl_100_127601_2008,jpcc_122_25668_2018,prb_101_155416_2020}, strategies to introduce them artificially are required in order to become more broadly applicable. Various different approaches have been reported in the literature, such as the use of electrical discharge,\cite{jmr_261_95_2015,ssnmr_100_26_2019,jmro_10_100043_2022} $\gamma$-irradiation,\cite{jpcl_10_4770_2019,ssnmr_119_101785_2022}, or UV irradiation.\cite{cej_23_8315_2017} The introduction of paramagnetic metal ions through doping is likely the most versatile and controlled approach and has been used in biological samples\cite{jbnmr_63_97_2015}, molecular crystals\cite{jacs_136_11716_2014} as well as in inorganic materials\cite{cpc_19_2139_2018} (Fig.~(\ref{fg:endoexo}~c)). Generally, a DNP experiment requires very low concentrations of polarizing agents, on the order of 10~mM,\cite{arpc_71_143_2020} and it is assumed that most of the structure will remain intact. Of course, this is an approximation, and it is known that in some cases the presence of even small amounts of imperfections can  alter the functional properties of a material.\cite{prm_2_125403_2018} An intriguing opportunity would be to exploit the dopants intended to boost sensitivity to characterize the structural modifications caused by their presence. However, the paramagnetic nature of the dopants pose some severe difficulties for such a quest.

The application of endogenous MI-DNP comes at the cost of increased synthetic demand and some lack of universality, as discussed in the previous subsection, but it offers some advantages. For instance, in samples with reactive surfaces, chemical interference will be reduced. And, most importantly, MI-DNP offers the ability to enhance the magnetization of otherwise inaccessible nuclei. This aspect will be treated in more depth in the next subsection.

    \subsection{Polarization vs. relaxation}
        \label{subsec:polvsRelax}
A fundamental question in any DNP experiment is to understand which fraction of the sample has been polarized. Knowledge of the spatial spread or localization of the hyperpolarization is required to be able to propose a quantitative interpretation of the NMR data. In DNP experiments the system is driven towards a non-equilibrium state, where hyperpolarization and longitudinal relaxation are competing. The degree of polarization of a nuclear spin will depend on the relative efficiency of both processes at its position. To avoid confusion, the time constant which describes how the system evolves towards this steady-state condition is termed the build-up time $T_{bu}$, as opposed to $T_1$. We will discuss two different scenarios: first we consider the case where spin diffusion does not contribute to the polarization transfer, when the nuclei are polarized directly by the polarizing agents. And second, the polarization of most nuclei will grow through the action of spin diffusion. We will term the two cases as $direct$ and $indirect$ DNP; note that this terminology has not been used consistently in the DNP literature, as some authors prefer to use the term indirect DNP only in cases where cross polarization is used. Furthermore, while the conceptual differentiation of the direct and indirect cases is clear, in most experiments it is likely that both effects contribute to the signal build-up to some extent.\cite{jacs_140_7946_2018}

    \subsubsection{Direct DNP}
First, we will consider the case where spin diffusion does not play a significant role. In this scenario, polarization only reaches a nucleus through direct polarization. Relaxation will occur either due to PRE from the polarizing agent or through an intrinsic relaxation mechanism, that it is independent of the polarizing agent. We have seen in subsection~\ref{subsec:SSpolarization} that the signal enhancement in the solid effect DNP mechanism depends on the saturation efficiency of the formally forbidden DQ or ZQ transition. In the appendix we show how the saturation efficiency depends on the various relaxation paths and on the effective irradiation frequency, in analogy to the Bloch equations. Combining equations~(\ref{eq:DNPsat}) and (\ref{eq:DNPenhancement}) we can write for the nuclear polarization:
        \begin{equation}
P_{n}\approx{P_{e}-\Delta p_{DQ}}={P_{e}-\frac{\Delta p_{DQ,eq}}{1+\tilde{\omega}_1^2\left[R_{2e}(2R_{1DQ}+2R_{1})\right]^{-1}}}.
\label{eq:Pn1}
\end{equation}
Looking at the terms in the denominator, both  $\tilde{\omega}_1^2$ and  $R_{1DQ}$ are proportional to ${\left(\omega_d^{en}\right)}^2$ (equations~(\ref{eq:effMW}) and (\ref{eq:R1dq})). Furthermore, if the nuclear relaxation time is governed by the paramagnetic relaxation enhancement, this also applies to the rate $R_{1}\propto{\left(\omega_d^{en}\right)}^2$ (equation~(\ref{eq:T1PRE_c})). Thus, the whole expression, and consequently the DNP signal enhancement, is independent of the strength of the dipolar coupling, and therefore, of the distance between the nucleus and the polarizing agent.  \cite{jpcl_11_5439_2020} 
A direct consequence of this distance independence is that as long as PRE dominates, the enhancements will be homogeneous and the DNP-NMR spectrum quantitative. This is mostly relevant for endogenous DNP, and in particular in MI-DNP in inorganic materials, where relaxation times in the absence of paramagnetic centres can be extremely long. Fig.~(\ref{fg:endoexo}~d) shows the results of numerical simulations mapping nuclear polarization as a function of the distance between electron and nucleus, confirming the analytical result.\cite{jpcl_11_5439_2020} 

The independence of the enhancement from the distance will no longer be warranted when the size of the PRE becomes comparable to the intrinsic relaxation rates. As can be seen in equation~(\ref{eq:Pn1}), if $R_{1}$ is independent of $\omega_d^{en}$, an increment in the distance will lead to a reduced signal enhancement. This relation also determines the penetration depth of hyperpolarization from exogenous polarizing agents in direct DNP to materials with low sensitivity nuclei (where spin diffusion is ineffective). Only if the longitudinal nuclear relaxation times are long are enhancements beyond the first surface layers obtained.\cite{ac_50_8367_2011,pccp_15_20706_2013,jacs_140_7946_2018,jpcl_10_3501_2019} The lack of penetration depth of the enhancement can be exploited to focus on surface-selective signal enhancements. The importance of this selectivity has received great attention. This approach for DNP is called DNP SENS (Surface Enhanced NMR Spectroscopy).\cite{jacs_132_15459_2010} The selectivity can be further increased by exploiting indirect DNP, where first the protons in the frozen solution containing the polarizing agent are hyperpolarized (assisted 
 by spin diffusion) and subsequently a heteronuclear cross polarization step transfers the polarization to the nuclei of interest at the material's surface.\cite{emr_7_93_2018,cocis_33_63_2018}\\

Another important implication of the distance independence is that the signal enhancement in MI-DNP  becomes independent of the concentration of paramagnetic dopants. This is also shown in Fig.~(\ref{fg:endoexo}~d), where the DNP signal enhancement of natural abundance $^{17}$O in Li$_4$Ti$_5$O$_{12}$ doped with variable amounts of Fe(III) was measured. The experimental results show a plateau of constant signal enhancement over almost one order of magnitude in Fe(III) mole fraction. Ultimately, the enhancement will decrease on both sides of the plateau. On the low concentration side, increasingly longer distances between polarizing agents will cause non-PRE mechanisms to dominate nuclear relaxation. This scenario will lead to inhomogeneous enhancements throughout the sample and might cause, for instance, a broadening of the lines observed under DNP, if the spins in the vicinity of the paramagnetic centres having larger enhancements are broadened due to paramagnetic effects.\cite{jacs_141_451_2019}

\begin{figure*}
\begin{center}
\includegraphics[scale=1]{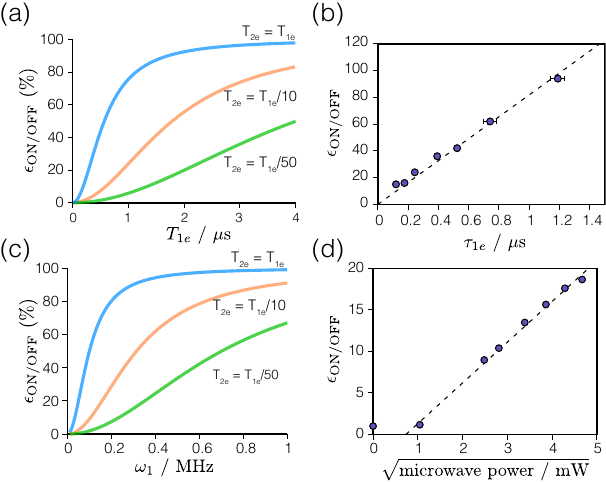}
\end{center}	
\caption{{\bf Dependence of the DNP enhancement on electron relaxation times and microwave irradiation power.} Steady-state nuclear polarization enhancements calculated (a and c) using equation~\ref{eq:Pn1} for an electron coupled to a nucleus with the gyromagnetic ratio of $^6$Li, in a magnetic field of 9.4~T and at 100~K. In (a) a fixed $\omega_1$ of 0.35~MHz was used,\cite{pccp_21_2166_2019} while in (b) a constant $T_{1e}$ of 2~$\mu$s was used. In addition, the value of $\omega_1$ in (a) and (c) was multiplied by a factor 3 to account for the effective central transition frequency of a spin 5/2, such as Mn(II).\cite{jcp_68_5518_1978} Experimental MI-DNP results show the $^6$Li signal enhancement in: (b) Li$_4$Ti$_5$O$_{12}$ doped with Fe(III) as a function of $\tau_{1e}$ (obtained according to equation~\ref{eq:T1efromT1T2} in variable temperature experiments). And (d) in Li$_4$Ti$_5$O$_{12}$ doped with Mn(II) as a function of the square root of the microwave power, while regulating the sample temperature to reduce artifacts. Figure (a) and (b) adapted with permission from Ref.~\cite{jmr_336_107143_2022}.}
\label{fg:MIDNPenh}
\end{figure*}

On the high concentration side of the enhancement plateau, strong electron couplings will reduce the  electron relaxation times (or more generally $\tau_{1e}$ and $\tau_{2e}$ as discussed in subsection~\ref{subsec:pre}), reducing the saturation efficiency. In cases where PRE is the dominant relaxation mechanism and assuming that the second part of the denominator in equation~(\ref{eq:Pn1}) is smaller than 1 (meaning that the enhancement, $\epsilon_{ON/OFF}$, is far from the theoretical maximum $\epsilon^{max}_{ON/OFF}$), one finds the following expression:
        \begin{equation}
P_n\propto\omega_1^2T_{1e}T_{2e}, \quad (\text{for $\epsilon_{ON/OFF}\ll\epsilon^{max}_{ON/OFF}$})
\label{eq:enhvsT1e}
\end{equation}
where we have taken $R_{1}\propto R_{1e}$ and $R_{1DQ}\propto R_{1e}$ (see equations~(\ref{eq:T1PRE_c}) and (\ref{eq:R1dq})). Large enhancements can be obtained when the electron relaxation times are long and/or the microwave power is large. The expected enhancements as a function of the relaxation times and the nutation frequency following these equations are shown in Fig.~\ref{fg:MIDNPenh} together with MI-DNP experimental results. Assuming that $T_{2e}$ is proportional to $T_{1e}$, then the enhancement has the same dependence on both  $\omega_1$ and $T_{1e}$. This is observed experimentally, where actually the enhancement appears to grow linearly with both (note that $\omega_1\propto\sqrt{P}$, where $P$ is the microwave power). We note that the theoretical expressions do not predict a linear behaviour, but as can be seen in Fig.~\ref{fg:MIDNPenh} (a and c) the curves follow a nearly linear behaviour over a large region around their inflection point.

    \subsubsection{Indirect DNP}
A different scenario is encountered in the presence of efficient spin diffusion. In this case, only nuclei in the vicinity of the polarizing agent are directly polarized, often referred to as core nuclei.\cite{jcp_134_074509_2011} Beyond these first layers of core nuclei an indirect transfer of polarization via spin diffusion becomes more efficient than the direct polarization step in spreading the polarization throughout the bulk of the sample. It is important to note that not all core nuclei will contribute to the polarization of the bulk due to a reduced efficiency of the spin diffusion rates in the immediate proximity of the paramagnetic agent. This effect is known as the spin diffusion barrier.\cite{pr_119_79_1960} As will be elaborated in the following, estimating the sizes of the core and bulk spin pools and of the spin diffusion barrier is not trivial.

Also in the indirect DNP case hyperpolarization is competing with nuclear relaxation, so that the degree of polarization will depend on the ratio between the rates of spin diffusion and relaxation. In spin diffusion, polarization is transferred between homonuclear coupled spins mediated by the flip-flop operators $\left(\hat{I}_+\hat{I}_-+\hat{I}_-\hat{I}_+\right)$.\cite{mssnap} At thermal equilibrium the density matrix commutes with the flip-flop operators and the magnetization is stationary. However, if the coupled spins deviate from their Boltzmann populations, the operator no longer commutes with the density operator and the polarization will evolve between both spins, while the total amount will be conserved. In a large ensemble of coupled spins with a polarization gradient or polarization heterogeneity, the effect of spin diffusion will be to homogenize the polarization throughout the ensemble. While highly complex, this effect is in principle reversible, as has been demonstrated experimentally.\cite{prl_69_2149_1992} From a macroscopic point of view this processes often resembles actual diffusion and therefore might be described heuristically by diffusion equations (as has long been described for spin diffusion in diamagnetic solids, see e.g. Chapter V, Section B in Ref.~\cite{ponm}).

In order to estimate a length scale that can be reached by hyperpolarization via spin diffusion while competing with the nuclear relaxation, the following equation has been proposed:\cite{jacs_128_10840_2006}
\begin{equation}
L=\sqrt{DT_1},
\end{equation}
where $L$ represents a characteristic distance from the polarizing agent up until which nuclear polarization will be enhanced and $D$ is the spin diffusion constant. The concentration of nuclear spins plays a fundamental role here; on the one hand, high concentrations are required for there to be a strong coupling network to allow a transfer of polarization by spin diffusion, while on the other hand, excessive concentrations can be detrimental as strong homonuclear couplings can reduce the nuclear relaxation times, which is particularly important for protons.\cite{jacs_126_10844_2004,ac_49_7803_2010,pccp_23_1006_2021}

Fig.~(\ref{fg:endoexo}~b) shows a typical polarization build-up behaviour from an exogenous polarization source in a protonated sample (with subsequent cross polarization to $^{13}$C for detection). The enhancement will be largest at the surface. In addition, since the presence of paramagnetic species shortens the relaxation time, signal enhancements will depend on the build-up time; as nuclei deeper into the bulk of the material will have longer relaxation times and smaller enhancements, the overall enhancement will decrease with increasing delays.\cite{jacs_134_16899_2012} If the particles are heterogeneous in composition, the relative intensities of the various NMR signals will not be quantitative, and this fact has actually been exploited to study sample homogeneity.\cite{mm_55_2952_2022} For protonated materials it has been shown that $^1$H spin diffusion can transport the polarization over micrometer length scales.\cite{jacs_128_10840_2006,jacs_134_16899_2012} For heteronuclei, the diffusion coefficients will be weaker since the homonuclear dipolar couplings $\omega_d^{nn}$ are smaller, nonetheless, in some cases with very long $T_1$ relaxation times, surprisingly long length scales of transfer have been reported.\cite{jacs_140_7946_2018,jmr_323_106888_2021} Of course, the same considerations are valid for endogenous DNP approaches like MI-DNP, with the difference that in this case the potential polarization gradients may be distributed throughout the bulk of the sample.

Assessing the diffusion constant $D$ in a quantitative manner can be particularly challenging. Abragam gives the following definition:\cite{ponm,mssnap}
\begin{equation}
D=Wa^2,
\end{equation}
where $a$ is the distance between neighbouring nuclei and $W$ the transition rate of the spin exchange. Obtaining the latter is not straightforward to determine but to a first approximation can be estimated by the strength of the dipolar couplings, which might be obtained from the width of the spectrum for a strongly coupled homonuclear spin bath. A more elaborate estimate can be obtained taking into account the intensity of the zero-quantum line shape at zero frequency $f(0)$ according to:\cite{ssnmroP_84_83_1998}
\begin{equation}
W=\frac{\pi}{2}\left(2B\omega_d^{nn}\right)^2f(0).
\end{equation}
Note that the zero-quantum line shape usually also depends on the interaction strength ($f(0)\propto\left(\omega_d^{nn}\right)^{-1}$), such that $W\propto\omega_d^{nn}$. To use this equation one would need to know the frequency difference as well as the width of the zero-quantum line shape ($T_{2,ZQ}$) for each relevant pair of spins in the sample.\cite{prb_32_5608_1985} The smaller the frequency mismatch the higher the efficiency, while the effect of the linewidth depends on the frequency difference. Large frequency shifts will occur at the proximity of the polarizing agents due to its paramagnetic nature. This can lead to a severe reduction of the transition rate between core and bulk nuclei: the spin diffusion barrier.\cite{pr_119_79_1960} It can play an important role for the enhancement factors and polarization build-up times.\cite{jcp_136_015101_2012} Assessing the size of the radius within the barrier (if it can be assumed to be spherical) as well as the degree of dampening of the polarization transfer efficiency across it, are extremely challenging quests. Further aggravated by the fact that nuclei within the proximity of paramagnetic centres are likely quenched.

The treatment of spin diffusion becomes even more complex if the effect of MAS has to be considered.\cite{jcs_84_3713_1988} On the one hand MAS will partially average out the dipolar couplings which will reduce the spin diffusion constant,\cite{jacs_128_10840_2006,jpcc_117_1375_2013,jacs_140_7946_2018} but on the other,  during `rotor events' MAS can enable energy matching conditions between spin pairs in the presence of large anisotropic interactions, eventually leading to an enhanced polarization transfer.\cite{cpl_71_148_1980} Recent advances in numerical simulations are providing helpful insights for understanding this complex problem in systems with increasing numbers of spins.\cite{prl_115_020404_2015,pccp_19_3506_2017,jpcl_11_5655_2020,jpc_156_124112_2022} This is particularly relevant in the context of the spin diffusion barrier. Recently various experimental approaches have tackled this problem, aiming to assess the spin diffusion barrier quantitatively by taking advantage of the DNP effect, either by estimating the distance between the electron spin and the most distant nuclei within the same molecule,\cite{sa_5_eaax2743_2019,chem_7_421_2021}  or, alternatively, by observing the signal evolve from a quenched but hyperpolarized state via spin diffusion towards the bulk of the sample.\cite{sa_7_eabf5735_2021}

An appealing approach for a simple estimation of macroscopic diffusion constants for different nuclei in various materials consists in scaling a known diffusion constant, such as that for $^{19}$F in CaF$_2$, according to:\cite{jacs_140_7946_2018}
\begin{equation}
D\propto c^{1/3}\gamma^2,
\end{equation}
where $c$ is the concentration of nuclei in the sample and the power of $1/3$ follows from geometric considerations, while the power of two on the gyromagnetic ratio originates from the linear dependence of $W$ on $\omega_d^{nn}$.

\section{Applications of metal ions for DNP}
    \label{sec:app}

    \subsection{The early days}
\begin{figure*}
\begin{center}
\includegraphics[scale=1]{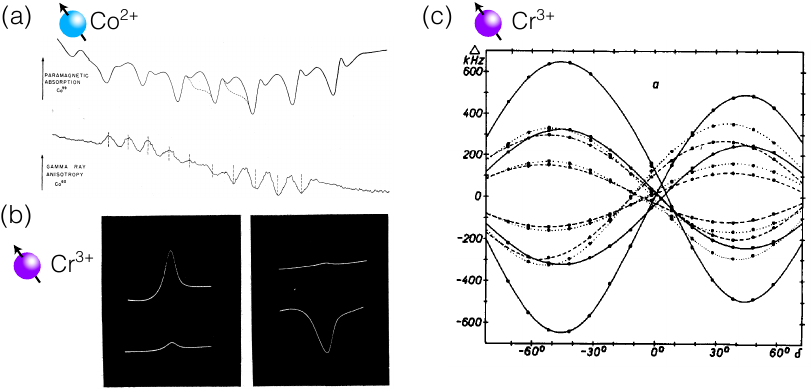}
\end{center}	
\caption{{\bf Early DNP measurements using paramagnetic metal ions as polarizing agents from 1957 (a), 1959 (b) and 1970 (c).} (a) EPR spectrum of Co(II) in La$_2$Mg$_3$(NO$_3$)$_{12}\cdot$24D$_2$O, with abundance ratios Mg:$^{59}$Co:$^{60}$Co of 10$^4$:50:1, measured at 1.6~K and 9.3~GHz, in the top figure. The 7/2 nuclear spin of $^{59}$Co gives rise to 8 main lines in the EPR spectrum, while the lines in between are attributed to forbidden transitions. The DNP effect is detected with $\gamma$-ray absorption anisotropy of the radioactive decay of hyperpolarized $^{60}$Co (bottom figure). Reproduced with permission from \cite{pr_106_165_1957}. (b) DNP enhancement of the $^{27}$Al NMR signal in a ruby crystal doped with 0.1\% Cr(III) through microwave irradiation at the DQ (left) and ZQ (right) frequencies, at 4.2~K and irradiation frequency of 9.3~GHz.Reproduced with permission from \cite{prl_3_13_1959}. (c) Frequency of $^{17}$O as a function of the angle between the c-axis of a Cr(III) doped Al$_2$O$_3$ crystal and the external magnetic field used to determine the $^{17}$O quadrupolar coupling constant at natural abundance. Acquisition of the signal was enabled through DNP at 1.9~K by constant microwave irradiation at 33~GHz and adjustment of the magnetic field at each angle for optimized signal enhancement.Reproduced with permission from \cite{pla_31_416_1970}.}
\label{fg:earlySE}
\end{figure*}

Some years after Carver's and Slichter's demonstration of DNP by the Overhauser mechanism in conducting materials,\cite{pr_92_411_1953,pr_92_212_1953}  a new mechanism of DNP in insulating materials, known as the solid effect,  was discovered.\cite{pr_106_164_1957,pr_106_165_1957,jpr_19_843_1958,prl_2_449_1959} During the second half of the 20$^{\text{th}}$ century various examples of the use of paramagnetic metal ions as polarizing agents for solid effect DNP were published (Fig.~\ref{fg:earlySE}). The enhancements were enabled by performing the measurements at very low temperatures (a few K), and low magnetic fields, using doped single crystals. These conditions facilitate saturation of the forbidden double- and zero-quantum transitions due to longer electron relaxation times at these temperatures and the larger effective nutation frequencies at low fields. In addition, inhomogeneous line broadening due to anisotropic interactions can be avoided in single crystals.

Initial experiments served to gain and confirm theoretical understanding of the solid effect mechanism.\cite{pr_122_1781_1961} As polarizing agents, Co(II), Cr(III), Ce(III) and iron impurities (presumably Fe(III)) were used.\cite{prl_2_449_1959} In the 1970's this approach was explored for characterizing NMR properties of challenging nuclei, such as $^{43}$Ca in CaF$_2$\cite{prl_32_1096_1974} and further to obtain the quadrupolar coupling parameter of $^{17}$O and $^{25}$Mg at natural abundance in Al$_2$O$_3$\cite{pla_31_416_1970} and Mg$_2$SiO$_4$,\cite{pla_66_150_1978} respectively.

    \subsection{Introducing metal ions for high-field MAS DNP in complexes}
\begin{figure*}
\begin{center}
\includegraphics[scale=1]{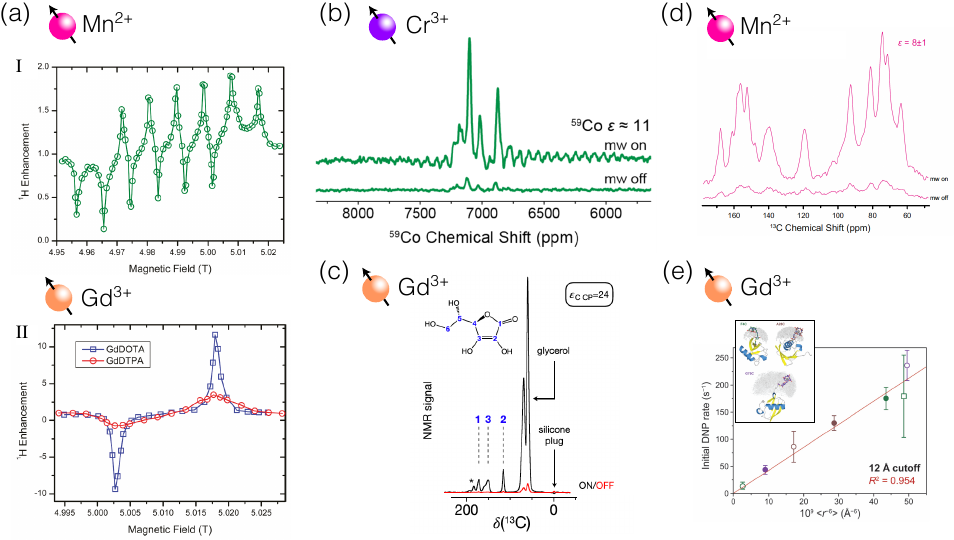}
\end{center}	
\caption{{\bf Metal ions in molecular complexes for DNP.} 
(a) First report of the use of paramagnetic metal ions as polarizing agents for DNP at high magnetic fields and temperatures in 2011 using MnDOTA, GdDOTA and GdDTPA. The $^1$H DNP sweep profiles obtained at about 85~K, a microwave frequency of 139.65~GHz and spinning at 5~kHz are shown in $I$ and $II$, for the Mn(II) and Gd(III) complexes, respectively. Reproduced with permission from Ref.~\cite{jacs_133_5648_2011} 
(b)Cr(III) was used as polarizing agent by doping into the molecular crystal [Co(en)$_3$Cl$_3$]2$\cdot$NaCl$\cdot$6H$_2$O. Shown is the $^{59}$Co DNP enhanced spectrum obtained with a doping level of 3\% Cr(III), at 85~K, 4~kHz MAS frequency and microwave frequency of 140~GHz. Reproduced with permission from  Ref.~\cite{jacs_136_11716_2014}. 
(c) In samples where nitroxide radicals are not stable, metal ions might still serve as polarizing agents. [Gd(tpatcn)] in a solution of asorbic acid was shown to yield significant enhancement, as shown in the $^{13}$C CPMAS spectrum, while in the same sample no enhancement is obtained with AMUPol. Experiments performed at 9.4~T, 100~K and 12.5~kHz MAS frequency. Reproduced with permission from  Ref.~\cite{jacs_141_8746_2019}.
(d) Site specific labeling of biomolecules with paramagnetic metal ions were  exploited to obtain $^{13}$C signal enhancements through direct polarization from Mn(II) in RNA at 9.4~T, approximately 100~K and MAS spinning at 10~kHz. Reproduced with permission from  Ref.~\cite{jbnmr_63_97_2015} 
(e) The $^{15}$N DNP polarization build-up rates in Gd(III) tagged biomolecules were used to accurately measure electron-nucleus distances at 9.4~T, 120~K and spinning at 8~kHz. Reproduced with permission from Ref.~\cite{pccp_22_25455_2020}.}
\label{fg:bioMIDNP}
\end{figure*}

Since the 1990's and 2000's DNP has experienced a revived interest, mostly lead by the advances introduced by the group of R. G. Griffin. The use of gyrotrons made it possible to obtain high power microwaves at the electron Larmor frequency at high magnetic fields,\cite{prl_71_3561_1993,jmr_160_85_2003,pccp_12_5850_2010} which enabled the use of DNP to gain sensitivity in high resolution MAS NMR experiments, initially using nitroxide mono- and biradicals.\cite{jacs_126_10844_2004} Subsequently, in 2011, the use of paramagnetic metal ions as polarizing agents at a magnetic field of 5~T, a temperature of 85~K and MAS frequencies of 5~kHz was demonstrated by Corzilius et al.\cite{jacs_133_5648_2011} (Fig.~\ref{fg:bioMIDNP}~a). In these first experiments, chelating ligands DOTA and DTPA where used to form complexes with high-spin Mn(II) and Gd(III) (spin 5/2 and 7/2, respectively). DNP experiments were performed by dissolving the complexes in d$_8$-glycerol/D$_2$O/H$_2$O (60:30:10~\%~v/v) at concentrations between 2 and 100~mM. Enhancements of up to a factor $\epsilon_{ON/OFF}=13$ were obtained for Gd-DOTA. Since this pioneering work, further developments in the structure of the chelating complexes  have lead to higher enhancements. For instance, it was shown that using Gd(III) complexes, enhancements correlate inversely with the square of the zero field interaction strength, leading to a higher enhancement factor in the complex Gd(tpatcn), specifically designed for this purpose  (Fig.~\ref{fg:bioMIDNP} c).\cite{jacs_141_8746_2019,jpcc_126_11310_2022} Bis(Gd-chelates) can be used to obtain enhancements via the cross effect DNP mechanism, which was shown to be particularly efficient for nuclei with low gyromagnetic ratios.\cite{ac_56_4295_2017} Alternatively, simpler formulations consisting in dissolving simple gadolinium salts, such as GdCl$_3$\cite{pccp_18_27205_2016} and Gd(NO$_3$)$_3$,\cite{jpcb_126_6281_2022} in organic glass-forming matrices have also been shown to provide routes for DNP, although the enhancement factors are lower.

In parallel, the use of metal ions as endogenous sources of polarization was also developed by Corzilius and co-workers, Firstly by introducing Cr(III) in the molecular crystal [Co(en)$_3$Cl$_3$]2$\cdot$NaCl$\cdot$6H$_2$O, where it replaces diamagnetic Co(III) (Fig.~\ref{fg:bioMIDNP}~b).\cite{jacs_136_11716_2014} 
DNP to $^1$H, $^{13}$C and $^{59}$Co was demonstrated, yielding low enhancements for $^1$H, but more than a factor of 25 for $^{59}$Co. For the latter two nuclei, the magnetic field sweep profile suggested the presence of cross effect DNP, although the source of the frequency difference between neighbouring electrons is not clear. Another example of endogenous DNP was achieved on an RNA molecule through selective binding of Mn(II) to the structure which then served as polarizing agent for $^{13}$C (Fig.~\ref{fg:bioMIDNP}~d).\cite{jbnmr_63_97_2015}

In biomolecules which cannot chelate metal ions, an alternative option is the use of site-specific tags. This was shown by the use various Gd-carrying tags bound to ubiquitin which were subsequently used to enhance the $^{13}$C NMR spectrum  via direct DNP.\cite{pccp_18_27205_2016} Interestingly, this work further showed strong evidence for the appearance of the cross effect DNP mechanism at large gadolinium concentrations, where the frequency difference between the coupled electrons was attributed to the ZFS of the central transition. 
The idea of DNP using tags was further developed for various site-specific tags to actually obtain distance constraitns, which can be related to structural arrangements of the molecule (Fig.~\ref{fg:bioMIDNP}~e).\cite{pccp_22_25455_2020}

    \subsection{Metal ions in inorganic oxides: The MI-DNP approach}
            \begin{figure*}
\begin{center}
\includegraphics[scale=1]{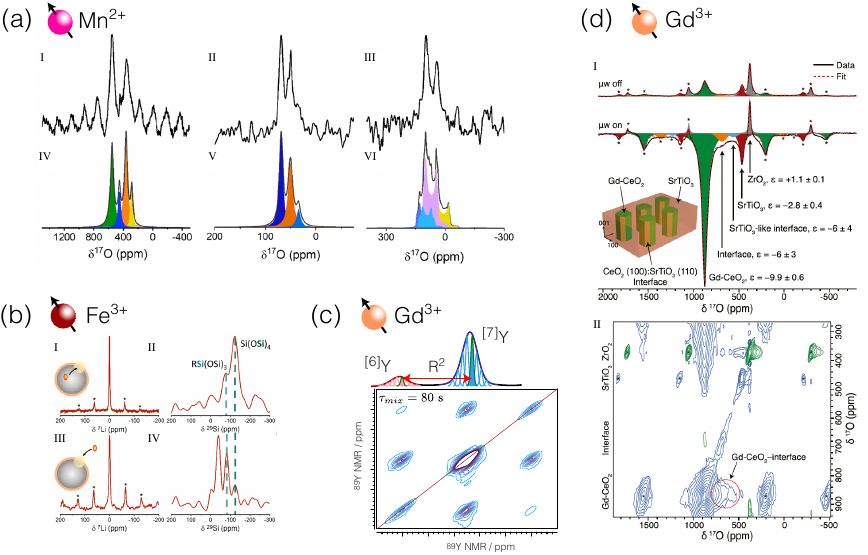}
\end{center}	
\caption{{\bf MI-DNP in inorganic oxides.} MI-DNP enhanced measurements performed at a magnetic field of 9.4~T, under MAS and at 100~K (a-c) or 140~K (d).
(a) $^{17}$O natural abundance MAS DNP spectra of ($I$) Li$_2$ZnTiO$_3$, ($II$) MgAl$_2$O$_4$ and ($III$) NaCaPO$_4$, with corresponding best fits below, all doped with Mn(II) as polarizing agent. Reproduced with permission from \cite{jacs_141_451_2019}.
(b) Comparison of the $^7$Li and $^{29}$Si MAS NMR spectra of an alkylated Li$_x$Si$_y$O$_z$ coating layer surrounding  TiO$_2$ particles obtained after DNP enhancement from MI-DNP from Fe(III) doped into the particle as endogenous polarizing agent (top) and from TEKPol in tetrachloroethane es exogenous polarizing agent. Reproduced with permission from \cite{jacs_143_4694_2021}. 
(c) 2D MAS DNP homonuclear $^{89}$Y correlation map in Ce$_{0.6}$Y$_{0.4}$O$_{1.8}$ doped with 0.1\% Gd(III). Reproduced with permission from \cite{jpcl_12_2964_2021} 
(d) One and two dimensional $^{17}$O MAS DNP NMR spectra of Gd(III) doped ceria nanopillars embedded in a SrTiO$_3$ matrix, with partial  $^{17}$O enrichment and spinning at 37~kHz MAS.Reproduced with permission from \cite{jpcc_125_18799_2021}.}
\label{fg:MIDNP}
\end{figure*}

In 2018 the authors' group showed that it is possible to obtain DNP enhancements using metal ions introduced  as dopants into the bulk of inorganic micron-sized particles, first under static conditions,\cite{cpc_19_2139_2018} and subsequently with MAS.\cite{jacs_141_451_2019} The first demonstrations of MI-DNP used Mn(II) and Gd(III) doped into Li$_4$Ti$_5$O$_{12}$ to enhance the $^7$Li NMR signal. Subsequently, in the MAS DNP experiments the enhancement obtained from Mn(II) metal ion dopants was exploited to measure $^{17}$O spectra at natural abundance for the battery anode materials Li$_4$Ti$_5$O$_{12}$ and  Li$_2$ZnTiO$_3$, as well as the phosphors  MgAl$_2$O$_4$ and NaCaPO$_4$ (Fig.~\ref{fg:MIDNP}~a). Somewhat larger enhancements of up to almost 300 were obtained by using Fe(III) as polarizing agent in the same Li$_4$Ti$_5$O$_{12}$ structure.\cite{jpcc_124_7082_2020,jpcl_11_5439_2020} More interestingly, the presence of Fe as a dopant in the material had a positive effect on its electrochemical performance and it was shown that it is possible to reversibly electrochemically  deactivate and reactivate the DNP performance of the Fe ions by conversion between Fe(II) and Fe(III). Other examples where MI-DNP was reported to give enhancements are the lithium ion conductor LiMgPO$_4$ doped with Mn(II),\cite{jmr_336_107143_2022} as well as a series of different oxide glasses doped with Gd(III) and Mn(II),\cite{jpcc_125_23126_2020}  although the enhancements obtained in the glasses are significantly smaller compared to crystalline materials. This discrepancy  was later attributed to intrinsically shorter $T_{1e}$ values in the glass as well as a significantly larger dielectric loss, based on a comparison between amorphous and crystalline calcium lithium silicates.\cite{jpcc_127_4759_2023}

A system which has been shown to yield very large DNP enhancements is gadolinium doped CeO$_2$. By introducing Gd(III) as polarizing agent in yttrium doped ceria, enhancements close to 200 were obtained for $^{89}$Y.\cite{jpcl_12_2964_2021} This gain in sensitivity enabled acquiring homonuclear 2D correlation spectra at the rotational resonance condition of $^{89}$Y, despite its low gyromagnetic ratio (Fig.~\ref{fg:MIDNP} c). These NMR results gave information regarding the medium-range distribution of oxygen vacancies in the sample, which is a fundamental property for its performance as oxygen ion conductor. Higher enhancements were obtained by doping pure ceria with Gd(III), up to a factor of 650 in $^{17}$O.\cite{jpcl_12_345_2020,jpcc_125_18799_2021} The difference in enhancements between ceria with and without yttrium arises from the oxygen vacancies introduced by yttrium, which can distort the otherwise very symmetric octahedral coordination environment of gadolinium, thus increasing its ZFS and likely shortening  its relaxation times.\cite{jcp_131_124515_2009} Both Gd(III) and Y(III) are aliovalent dopants, replacing Ce(IV) and thus creating one half of a vacancy per substitution. Thus, this is a nice example of how the vacancies introduced upon doping are not necessarily adjacent to the dopant, as otherwise the average Gd(III) environment would not show a loss in symmetry.

A different application of MI-DNP is the study of heterogeneous materials. By doping one specific phase of the material with a paramagnetic polarizing agent it is possible to transfer the polarization to adjacent phases, so as to give direct information on the interfacial structure or composition. Two such applications are shown in Fig.~\ref{fg:MIDNP} (b) and (d). In the work by Haber et al,\cite{jacs_143_4694_2021}  TiO$_2$ particles were doped with Fe(III) and subsequently coated with an alkylated lithium silicate layer. The MI-DNP enhanced $^{29}$Si spectrum showed a markedly different composition compared to a $^{29}$Si spectrum obtained with DNP using an exogenous nitroxide biradical. These results proved that the composition was not homogenous across the coating layer. In the second example, by Hope et al.\cite{jpcc_125_18799_2021} ceria nanopillars doped with Gd(III) were grown into a SrTiO$_3$ structure. The obtained DNP enhanced spectrum highlighted the interfacial environments.

    \subsection{Future perspectives}
 
\begin{figure*}
\begin{center}
\includegraphics[scale=1]{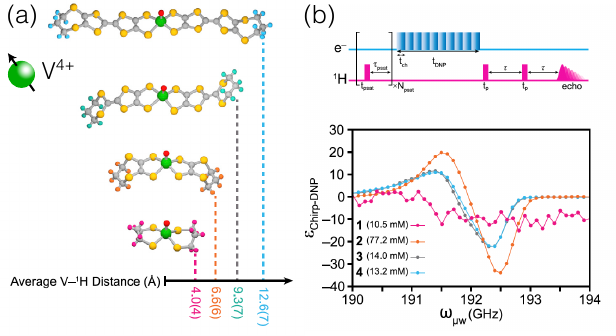}
\end{center}	
\caption{{\bf Further developments in microwave capabilities will increase the pool of metal ions for MI-DNP.} (a)~Vanadyl complexes with controlled V(IV)-H distances.  (b) DNP frequency sweep profiles for the compounds shown in (a) (the numbers shown in (b) correspond to the structures in order of increasing length, 1 being the shortest and 4 the longest). Experiments performed at 4~K and 6.9~T using chirped microwave pulses, microwave irradiation  was applied only during the build-up time. Reproduced with permission from \cite{chem_7_421_2021}.}
\label{fg:Vanadium}
\end{figure*}

Over recent years, DNP instrumentation has seen great advances. It is clear that MI-DNP would benefit significantly from further improvements in DNP instrumentation. Currently, in most known MI-DNP applications the signal enhancements are limited by the difficulty of saturating efficiently the formally forbidden double and zero quantum transitions. This is due to the short electron relaxation times of the metal ions compared to organic radicals, as well as to the higher power requirements of the solid effect compared to the cross effect. Thus, enhancements in MI-DNP should grow significantly when higher power microwave sources become available, or microwave cavities are introduced.\cite{jmr_223_170_2012,emagres_7_179_2018} In addition, currently available commercial instrumentation has a relatively narrow range of field sweep. Minor deviations from $g_e=2$ can make metal ions inaccessible to current instruments. This is for instance the case for the metal ions Cr(III) and V(IV), which have been proven useful as polarizing agents with home-built instrumentation at intermediate magnetic fields (Fig~\ref{fg:Vanadium}).\cite{jacs_136_11716_2014,chem_7_421_2021} The reported values of $g_e$, however, deviate slightly from 2, due to weak spin-orbit couplings, a deviation that  hinder their access using currently available commercial instruments. 

As hopefully has become clear from this review, development of MI-DNP requires strong integration of EPR and NMR results. Therefore, we expect significant advances unfolding further capabilities of MI-DNP from the implementation of sophisticated pulsed EPR methods to this purpose. In addition, numerical DNP simulations are becoming more efficient, allowing one to increase the size of the spin systems and to include high-spin species.\cite{jmr_333_107106_2021} A deeper understanding of the  spin dynamics involved should help to define optimum conditions for MI-DNP.

As for specific applications to materials science, the possibilities seem endless, and we expect an increasing number of material classes benefiting from MI-DNP.  Very recently, the possibility of applications beyond oxides was shown to be feasible in the Mn(II) doped perovskite  Cs$_2$NaBiCl$_6$.\cite{jacs_145_4485_2023} The reported enhancements are very low under MAS conditions at 100~K, however, they were shown to be larger at 20~K, highlighting new opportunities of MI-DNP as instrumentation is further improving. Further, incorporation of metal ion dopants into specific phases of the material can help in understanding heterogeneous materials beyond the reach of exogenous DNP formulations,\cite{jacs_143_4694_2021}  which are mostly limited to the outermost layers of a sample. Analysis of the initial polarization build-up rates has the potential to map the architecture of interfacial regions layer-by-layer.\cite{pccp_22_25455_2020}

\section{Conclusions}
We have presented here a review of the theoretical concepts and recent developments of DNP based on metal ions. A growing increase in understanding of the magnetic resonance properties involved in the MI-DNP process will facilitate its implementation towards new classes of materials. MI-DNP possesses some fundamental differences compared to exogenous MAS DNP, making it a unique tool for boosting the NMR sensitivity of nuclear spins in the bulk of materials. Therefore, we believe that it has the potential to become a significant complement to solid state NMR characterization of materials on the one hand, and on the other that it can be tailored to address specific challenging questions of structural heterogeneity, mainly for functional materials.  

The main challenges that remain include simple $a priori$ identification of promising strategies for sample preparation for new classes of materials, characterization of the magnetic resonance properties of the paramagnetic dopants and, finally, optimization of favourable DNP conditions. However, with the increasing number of known protocols and a deeper understanding of the underlying spin physics, more tools for solving these challenges are emerging and should become more accessible to intuition.

\section{Acknowledgements}
We would like to thank Adi Harchol and Tamar Wolf for sharing the data which was used in Figs.~(\ref{fg:spindiff}~b) and (\ref{fg:MIDNPenh}~d), respectively, and Yonatan Hovav and Daphna Shimon for sharing with us their DNP simulation codes which were used to create Figs.~(\ref{fg:enhpop}~b) and (\ref{fg:sweeps}~a) and which have been an invaluable help for our journey in the field of DNP. This work was funded by the European Research Council (MIDNP, Grant 803024).  The work was made possible in part by the historic generosity of the Harold Perlman family.

\appendix
    \section{Relaxation and saturation}
        \subsection{The Bloch equations}

       \begin{figure}
\begin{center}
\includegraphics[scale=1]{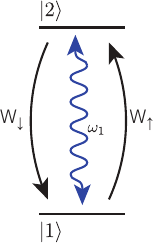}
\end{center}	
\caption{{\bf Saturation vs. relaxation.} Energy level diagram of a single spin 1/2 showing the direction of the transition probabilities $W_{\uparrow}$ and $W_{\downarrow}$. The wavy arrow represents an external irradiation of the transition with the amplitude given by the nutation frequency $\omega_1$.}
\label{fg:saturation}
\end{figure}

In magnetic resonance we differentiate between two relaxation processes, namely longitudinal and transverse relaxation. Longitudinal relaxation refers to the process that brings the populations of the energy levels to their equilibrium value. Generally, the population difference between different energy levels is determined by Boltzmann statistics, which depends on the energy difference between the states and the temperature of the system. The time evolution of the magnetization vector ${\bf M}$ in the rotating frame is described by the Bloch equations of relaxation:\cite{pr_70_460_1946}
\begin{equation}
\frac{d}{dt}{\bf M}(t)=-{\bf R}({\bf M}(t)-{\bf M}_{eq}),
\end{equation}
with 
\begin{equation}
{\bf M}=
\begin{pmatrix}
M_x \\
M_y \\
M_z   
\end{pmatrix}
,\quad
{\bf M}_{eq}=
\begin{pmatrix}
0 \\
0 \\
M_{eq}
\end{pmatrix}
\quad \text{and} \quad 
{\bf R}=
\begin{pmatrix}
R_2 &   0   &   0\\
0   &   R_2 &   0\\
0   &   0   &   R_1
\end{pmatrix}
,
\end{equation}
where ${\bf R}$ is the relaxation matrix. It follows for the longitudinal and transverse magnetization, respectively:
\begin{equation}
\frac{d M_z(t)}{dt}=-R_1(M_z(t)-M_{eq}),
\label{eq:BlochT1}
\end{equation}
and
\begin{equation}
\frac{d M_{x,y}(t)}{dt}=-R_2 M_{x,y}.
\label{eq:BlochT2}
\end{equation}

The solutions to equations~(\ref{eq:BlochT1}) and (\ref{eq:BlochT2}) are:
\begin{equation}
 M_z(t)=M_{eq}-(M_{eq}-M_z(0))\exp[-(tR_{1})],
\label{eq:T1}
\end{equation}
and
\begin{equation}
M_{x,y}(t)= M_{x,y}(0)\exp[-(tR_{2})].
\label{eq:T2decay}
\end{equation}
In a saturation recovery experiment $M_z(0)=0$ and equation~(\ref{eq:T1}) simply becomes:
\begin{equation}
M_z(t)=M_{eq}[1-\exp[-(tR_{1})]].
\label{eq:T1satrec}
\end{equation}
 
The system approaches thermal equilibrium following an exponential behaviour characterized by the longitudinal relaxation rate ${R}_1=1/{T}_1$, while the transverse magnetization decays exponentially with the transverse relaxation rate ${R}_2=1/{T}_2$. 

        \subsection{The relaxation superoperator}
In order to analyze a DNP experiment it will be convenient to describe relaxation in Liouville space, which can capture the evolution of individual terms of the spin density matrix, even in a coupled spin system, in terms of transition probabilities. First we will derive the relaxation equations in this formalism for the most simple case, a single spin 1/2. Subsequently, we will introduce the effect of continuous irradiation. These concepts will then be expanded to the relevant coupled two spin system under continuous irradiation which can give rise to the DNP effect.

We start by transforming the density matrix from Hilbert to Liouville space by rearranging the matrix into a column vector according to:\cite{ponmr12d}
\begin{equation}
\tilde{\hat{\rho}}=
\begin{pmatrix}
p_1 &   c_{12}   \\
c_{21}  &   p_2 
\end{pmatrix}
\quad \longrightarrow \quad
\tilde{\hat{\rho}}=
\begin{pmatrix}
p_1  \\  
c_{21} \\
c_{12}   \\
p_2
\end{pmatrix}
.
\label{eq:rhoHiltoLiou}
\end{equation}
We now can look at the equation of motion, describing the time derivative of the density matrix in the rotating frame:
\begin{equation}
\frac{d\tilde{\hat{\rho}}(t)}{dt}=-\doublehat{R}\tilde{\hat{\rho}}(t)=
-
\begin{pmatrix}
W_{\uparrow} &   0    &0  &-W_{\downarrow}    \\
0   &R_2  &0 &   0  \\
0 &   0    &R_2 &0 \\
-W_{\uparrow} &      &0  &W_{\downarrow} 
\end{pmatrix}
\begin{pmatrix}
p_1 (t) \\  
c_{21}(t) \\
c_{12}(t)   \\
p_2(t)
\end{pmatrix}
,
\end{equation}
where $\doublehat{R}$ is the relaxation superoperator and $W_{\uparrow}$ and $W_{\downarrow}$ are the transition probabilities per second, shown in Fig.~(\ref{fg:saturation}).\cite{pomr} Note that the transition probabilities act on the populations in complete analogy to an exchange matrix on two exchanging pools in the Bloch-McConnell equations. From this equation we can write the evolution of the populations in terms of coupled differential equations:
\begin{equation}
\begin{split}
\frac{d p_1(t)}{dt}&=-p_1W_{\uparrow}+p_2W_{\downarrow},
\\
\frac{d p_2(t)}{dt}&=-p_2W_{\downarrow}+p_1W_{\uparrow}.
\end{split}
\end{equation}
The magnetization is given by the difference in population between both states, ${\Delta p=p_1-p_2}$.
\begin{equation}
\frac{d \Delta p(t)}{dt}=-2p_1W_{\uparrow}+2p_2W_{\downarrow}.
\end{equation}
From this equations, one can show by introducing the definitions\cite{pomr} ${p_{eq}=(p_1+p_2)\frac{W_{\downarrow}-W_{\uparrow}}{W_{\downarrow}+W_{\uparrow}}}$ and ${R_1=W_{\downarrow}+W_{\uparrow}}$ that:
\begin{equation}
\frac{d \Delta p(t)}{dt}=-R_1 (\Delta p(t)-p_{eq}),
\end{equation}
which is equivalent to equation~(\ref{eq:BlochT1}).

To ensure that the equations predict a Boltzmann distribution at thermal equilibrium the transition probabilities  $W_{\uparrow}$ and $W_{\downarrow}$ are defined from the equilibrium equation:
\begin{equation}
W_{\downarrow}=\frac{p_1}{p_2}W_{\uparrow}=\exp\left(-\frac{E_1-E_2}{k_BT}\right)W_{\uparrow}=\epsilon_{12}W_{\uparrow},
\end{equation}
leading to:
\begin{equation}
\begin{split}
W_{\uparrow}&=\frac{1}{1+\epsilon_{12}}R_1
, \\
W_{\downarrow}&=\frac{\epsilon_{12}}{1+\epsilon_{12}}R_1.
\end{split}
\end{equation}
        \subsection{Saturation under continuous irradiation}
Next, we will consider the effect of continuous application of an external oscillating magnetic field with amplitude $\omega_1$, assumed to be perfectly on resonance with the $p_1\leftrightarrow p_2$ transition. The matrix representation of the Hamiltonian in the rotating frame in Hilbert and Liouville space is:\cite{ponmr12d}
 
 \begin{equation}
\tilde{\hat{H}}_1=\omega_1\hat{I}_x=
\frac{1}{2}
\begin{pmatrix}
0 &  \omega_1    \\
\omega_1   &   0 
\end{pmatrix}
\quad \longrightarrow \quad
\tilde{\doublehat{H}}_1=
\frac{1}{2}
\begin{pmatrix}
0 & \omega_1    &-\omega_1  &0    \\
\omega_1   &0  &0 &  -\omega_1  \\
-\omega_1 &   0    &0 &\omega_1 \\
0& - \omega_1    &\omega_1  &0
\end{pmatrix}
.
\end{equation}
 The equation of motion considering both relaxation and the effect of the oscillating field is given by:
\begin{equation}
\frac{d\tilde{\hat{\rho}}}{dt}=\left(-i\tilde{\doublehat{H}}_1-\doublehat{R}\right)\tilde{\hat{\rho}}(t)
.
\end{equation}
Again, we obtain the longitudinal magnetization from the difference between populations:
\begin{equation}
 \begin{split}
\frac{d \Delta p(t)}{dt}&=-2p_1W_{\uparrow}+2p_2W_{\downarrow}+ic_{21}\omega_1-ic_{12}\omega_1\\
&=-R_1 (\Delta p(t)-p_{eq})-i\omega_1(c_{12}-c_{21}),
\end{split}
\end{equation}
further, it will be convenient to look at the difference between the coherences:
\begin{equation}
 \begin{split}
\frac{d \left(c_{12}-c_{21}\right)}{dt}&=ip_1\omega_1-ip_2\omega_1-c_{21}R_2+c_{12}R_2\\
&=i\Delta p(t)\omega_1+R_2\left(c_{12}-c_{21}\right).
\end{split}
\end{equation}
At steady state, we can combine both equations to obtain the well-known Bloch equation for saturation:\cite{pr_70_460_1946}
\begin{equation}
\Delta p=\frac{p_{eq}}{1+\omega_1^2T_1T_2}.
\end{equation}

         \subsection{Expansion to a two spin system}
Analogously one can derive the steady-state equations for a two spin system, as depicted in Fig.~\ref{fg:enhpop}. Assuming a pulse along $x$ perfectly on resonance with the DQ transition ($\bra{}{2}\leftrightarrow\bra{}{3}$), the relevant equation of motion is:   
\begin{equation} 
\frac{d\hat{\rho}(t)}{dt}
=-
\begin{pmatrix}
W_{\uparrow}^{n}+W_{\uparrow}^{e}+W_{\uparrow}^{ZQ} &  -W_{\downarrow}^{n}   &   -W_{\downarrow}^{e}   &-W_{\downarrow}^{ZQ}  &0  &0\\

-W_{\uparrow}^{n}   &W_{\downarrow}^{n}+W_{\uparrow}^{DQ}+W_{\uparrow}^{e} &  -W_{\uparrow}^{DQ}   &   -W_{\downarrow}^{e}   &-i\tilde{\omega}_1/2  &i\tilde{\omega}_1/2\\

 -W_{\downarrow}^{e}   &   -W_{\downarrow}^{DQ}   &W_{\uparrow}^{ZQ}+W_{\uparrow}^{n}+W_{\uparrow}^{e} & -W_{\downarrow}^{n}  &i\tilde{\omega}_1/2  &-i\tilde{\omega}_1/2 \\
 
 - W_{\downarrow}^{ZQ}   &  - W_{\downarrow}^{e}   &-W_{\downarrow}^{n} & W_{\uparrow}^{ZQ}+W_{\uparrow}^{n}+W_{\uparrow}^{e} &0  &0\\
0 & -i\tilde{\omega}_1/2   &  i\tilde{\omega}_1/2  &0  &R_{2DQ}  &0\\
0 & i\tilde{\omega}_1/2   &  - i\tilde{\omega}_1/2   &0  &0  &R_{2DQ}
\end{pmatrix}
\begin{pmatrix}
p_1\\
p_2\\
p_3\\
p_4\\
c_{32}\\
c_{23}
\end{pmatrix}
.
\end{equation}         

From here one can proceed following the same steps as in the single spin 1/2 case. Of course, it will be a more tedious and lengthy calculation, and here we will refer to the SI of Reference~\cite{jpcl_11_5439_2020} for a step-by-step analysis. It can be shown that the saturation efficiency for irradiation on-resonance at the DQ transition is given by:
\begin{equation}
\Delta p_{DQ}=\frac{\Delta p_{DQ,eq}}{1+\tilde{\omega}_1^2\left[R_{2e}(2R_{1DQ}+2R_{1})\right]^{-1}}.
\end{equation}
To obtain this equation, some (reasonable) approximations are required: $1)$ the electron relaxation rate is much faster than any other process, such that the ratio between populations connected by $R_{1e}$ remains constant. $2)$ The ZQ and DQ relaxation rates are equal: $R_{DQ}\approx R_{ZQ}$. $3)$ The difference in the nuclear transition rates is small compared to other rates: $W_{\uparrow}^{n}\approx W_{\downarrow}^{n}$. And finally, $4)$ The ratio in population between two energy levels separated by the electron Zeeman energy is close to unity (high temperature approximation).

    \section{Sample preparation and characterization}

        \subsection{Chemical doping}

Generally in MI-DNP the paramagnetic metal ion is intended as a structural spy in the host lattice, therefore, it is desired that the dopant not only enters the structure, but that its presence disturbs the materials original structure as little as possible.\cite{ssca} A strategy towards this goal consists in substituting stoichiometric amounts of a given cation in the structure by the paramagnetic agent. Ideally, both cations should have the same charge and similar ionic radii\cite{acb_25_925_1969} to avoid introducing vacancies or large strains (a variation of about 15\% in radii has been proposed as rough limit for formation of an extensive solid solution\cite{ssca}). In practice, common dopant concentrations used in MI-DNP are very small and defects caused by their presence likely negligible with respect to other sources of stoichiometric imbalance.  To ensure stoichiometric substitution, the metal ion precursor should be added prior to the material's synthesis.

The pool of known metal ions that have a potential as polarizing agents is limited, thus, it is not always possible to find a dopant fulfilling the above conditions. Nonetheless, it still might be possible to introduce a suitable paramagnetic metal ion into the structure as a dopant without creating a new phase. In such cases, it might not be obvious where the dopant will be located, not even whether it will occupy a substitutional or an interstitial site. Due to the very low concentrations, characterization of the dopant with conventional diffraction methods might not be possible. Instead, EPR might offer the most direct measure of the dopants environment.\\

A very important parameter in the MI-DNP experiment is the concentration of the paramagnetic dopant. For DNP purposes, the relevant parameter is the molar concentration, as this determines the radius of influence, as well as the mean distance between paramagnetic centres. The molar concentration  $c$ is given in moles per liter and can directly be obtained from the stoichiometric ratio $x$ for a known crystal structure:
\begin{equation}
c=\frac{xn_{cell}}{V_{cell}N_A}.
\end{equation}
Where $n_{cell}$ and $V_{cell}$ are the number of formula units per unit cell and the unit cell volume, respectively, and $N_A$ is Avogadro's number.

Alternatively, the molar concentration can also be calculated from the materials density $\rho_{batch}$ and molar mass $M_{batch}$, this might be required for non-crystalline materials:
\begin{equation}
c=\frac{\rho_{batch}}{M_{batch}}\frac{x}{1-x}.
\end{equation}

Assuming a homogenous distribution of paramagnetic centres, one can estimate the mean distance between paramagnetic dopants from twice their Wigner-Seitz radius.
\begin{equation}
r_{WS}=\left(\frac{3}{4\pi \rho_n}\right)^{1/3}.
\end{equation}
with $\rho_n$, the number density per unit volume:
\begin{equation}
\rho_{n}=\frac{N}{V}.
\end{equation}

        \subsection{Doping homogeneity}
It does not suffice that the dopant is introduced into the lattice of the material of interest; it is further critical to ensure a $homogeneous$ distribution of the polarizing agents throughout the targeted phase. This will ensure a more homogeneous enhancement and reduce interactions among paramagnetic centres that could lead to reduced relaxation times. in the case of solid state synthesis approaches, the homogeneity might be improved during sample preparation by thorough grinding and mixing of the precursors, longer sintering times or higher temperatures. 

Another challenge consists in actually determining whether the dopants are distributed in a homogeneous fashion throughout the sample. Again, X-ray diffraction methods are limited because of the low concentration of the polarizing agent. On the other hand, microscopy based analysis, might be limited to the surface or lack depth resolution, which could yield dopant distributions which are not representative of the bulk of the material. A powerful method (and convenient in the context of MI-DNP) for characterizing the dopant homogeneity is NMR. Mapping of the nuclear relaxation times\cite{prl_65_614_1990,prb_50_822_1994,pccp_16_18788_2014,pccp_19_12175_2017} or quenching of the signal\cite{pccp_18_9752_2016} as a function of the dopant concentration has been shown to provide a good indication of the doping homogeneity. Care should be taken at higher dopant contents, as it is known that the electron relaxation times can shorten due to electron-electron interactions, thus altering their effect on the nuclear spins.\cite{jmr_336_107143_2022} This is particularly important for slowly relaxing paramagnetic centre, as the ones relevant for MI-DNP.

    \section{Quantifying sensitivity gains}
Ultimately, the main objective of applying DNP in chemistry or materials research is to increase the signal sensitivity. Therefore, it would be useful to have a metric to compare the merit to sensitivity gain between various experimental approaches. This, however, turns out to be a surprisingly challenging task. Of course, the simplest measure is a comparison in signal intensity between experiments with and without microwave irradiation: $\epsilon_{ON/OFF}$. While this is a good metric for the efficiency of the DNP processes, it is known to have flaws when it comes to quantifying absolute sensitivity gains.\cite{ssnmr_66_6_2015,emr_105_2018} The issues of depolarization have been extensively discussed in the literature,\cite{jcp_140_184201_2014,pccp_17_21824_2015}, however, since the cross effect does not seem to be efficient in MI-DNP, this effect is likely not very relevant for MI-DNP. Probably most important for MI-DNP, and endogenous DNP in general, is the effect of signal quenching.\cite{jmr_240_113_2014,jmr_240_113_2014,pccp_21_10185_2019,jmr_336_107143_2022} A higher content of metal ion dopants can lead to larger values of $\epsilon_{ON/OFF}$ but actually lower the signal-to-noise ratio per scan.\cite{jpcc_124_7082_2020,jpcl_11_5439_2020} At the same time, higher paramagnetic content reduces the $T_1$ relaxation times and thus allows one to shorten the recycle delays. Besides quenching, an additional negative effect due to the shortening of $T_2$ relaxation times is a reduction in efficiency of CPMG detection schemes\cite{cs_3_108_2012},  and a possible broadening of the NMR resonances, eventually at the expense of resolution. Other variables that have non-trivial effects on the sensitivity of an experiment under DNP as they affect various processes in opposing ways are the temperature and the size of the external magnetic field. 

At this point we might draw an analogy to assessing the advantages of different rotor sizes for MAS NMR, where the right choice often requires prior knowledge about the sample studied. Therefore, for an unknown sample, determining the $best$ strategy for optimum signal sensitivity is not possible $a priori$. We hope, however, that this review will help spectroscopists to appreciate the importance of the relevant factors for the outcome of a MI-DNP experiment, increasing the odds of choosing a $good$ strategy within the given possibilities, and in this way enabling measurements with sufficient sensitivity.

\bibliography{Biblio-Bibtex}

    \section*{Glossary of abbreviations}
$aMAT$: adiabatic magic angle turning 

$bCTbK$: bis-cyclohexyl-TEMPO-bisketal

$DQ$: double quantum

$CE$: cross effect

$CPMG$: Carr Purcell Meiboom Gill

$CT$: central transition

$CW$: continuous wave

$DNP$: dynamic nuclear polarization

$DOTA$: 1,4,7,10-tetraazacyclododecane-1,4,7,10-tetraacetic acid

$DTPA$: diethylenetriaminepentaacetic acid

$EPR$: electron paramagnetic resonance

$FC$: Fermi contact

$HF$: hyperfine

$MAS$: magic angle spinning

$MI$-$DNP$: metal ion based dynamic nuclear polarization

$NMR$: nuclear magnetic resonance

$OE$: Overhauser effect

$PRE$: paramagnetic relaxation enhancement

$SE$: solid effect

$ST$: satellite transition

$tpatcn$: 1,4,7-tris[(6-carboxypyridin-2-yl)methyl]-1,4,7-triazacyclononane)

$TCE$: tetrachloroethane

$ZFS$: zero-field splitting

$ZQ$: zero quantum

$\mu W$: microwave

\end{document}